\documentclass[12pt]{article}
\usepackage{epsfig}
\begin{document}
\title{Momentum dependence of near-threshold photoproduction of $\Xi^-$ hyperons
 off nuclei and their properties in the nuclear medium}
\author{E. Ya. Paryev$^{1,2}$\\
{\it $^1$Institute for Nuclear Research, Russian Academy of Sciences,}\\
{\it Moscow 117312, Russia}\\
{\it $^2$Institute for Theoretical and Experimental Physics,}\\
{\it Moscow 117218, Russia}}

\renewcommand{\today}{}
\maketitle

\begin{abstract}
We study the near-threshold inclusive photoproduction of $\Xi^-$ hyperons
off $^{12}$C and $^{184}$W target nuclei within a first-collision model relying on the nuclear spectral function
and including incoherent $\Xi^-$ production in direct elementary ${\gamma}p \to K^+{K^+}\Xi^-$ and
${\gamma}n \to K^+{K^0}\Xi^-$ processes. The model takes into account
the impact of the nuclear effective scalar $K^+$, $K^0$, $\Xi^-$ and their Coulomb potentials on these
processes as well as the absorption of final $\Xi^-$ hyperons in nuclear matter, the binding of intranuclear
nucleons and their Fermi motion. We calculate the absolute differential cross sections and their ratios for the production of $\Xi^-$ hyperons off these nuclei at laboratory polar angles $\le$ 45$^{\circ}$ by photons
with energies of 2.5 and 3.0 GeV, with and without imposing the phase space constraints
on the $\Xi^-$ emission angle in the respective ${\gamma}p$ center-of-mass system.
We also calculate the momentum dependence of the transparency ratio for $\Xi^-$ hyperons for the
$^{184}$W/$^{12}$C combination at these photon energies.
We show that the $\Xi^-$ momentum distributions (absolute and relative)
at the adopted initial photon energies possess a definite sensitivity to the considered changes
in the scalar $\Xi^-$ nuclear mean-field potential at saturation density $\rho_0$ in the low-momentum range
of 0.1--0.8 GeV/c. This would permit evaluating the $\Xi^-$ potential in this momentum range
experimentally. We also demonstrate that the momentum dependence of the transparency ratio for $\Xi^-$
hyperons for both photon energies can hardly be used to determine this potential reliably.
Therefore, the precise measurements of the differential cross sections (absolute and relative)
for inclusive $\Xi^-$ hyperon photoproduction on nuclei near threshold in a dedicated experiment at
the CEBAF facility will provide valuable information on the $\Xi^-$ in-medium properties,
which will be supplementary to that inferred from studying of the ($K^-$,$K^+$) reactions
at initial momenta of 1.6--1.8 GeV/c.
\end{abstract}

\newpage

\section*{1 Introduction}

\hspace{0.5cm} The investigation of the ${\Xi^-}N$ and ${\Xi^-}$--nucleus interactions has received
considerable interest in recent years (see, for example, [1]). The knowledge of these interactions is important
for understanding the neutron star properties and the properties of such exotic nuclear systems as the $\Xi^-$ hypernuclei [1] as well as for constraining the equation of state (EoS) of dense nuclear matter [2].
Mostly, the study of $\Xi^-$ hypernuclei provide a valuable information on them at low energies.
Presently, the possibility to extract information on the ${\Xi^-}N$ interaction also from studying
the momentum correlations of $p$ and $\Xi^-$, produced in high-energy heavy ion collisions, as well as from
lattice QCD calculations becomes feasible (see, for instance, [3]).
An information inferred in Refs. [4] and [5--9] from old and new emulsion hypernuclear data, from missing-mass measurements [10--14] in the inclusive ($K^-$,$K^+$) reactions on nuclear targets at incident momenta
of 1.6--1.8 GeV/c, from the analysis the results of these measurements in the $\Xi^-$ bound and quasi-free regions
in Refs. [15--17] using the DWIA method as well as the theoretical predictions [18--31]
indicate that the $\Xi^-$ hyperon feels only a moderately attractive low-energy potential in nuclei,
the depth of which is about of -(10--20) MeV at saturation nuclear density $\rho_0$.
To improve significantly our knowledge on the low-energy $\Xi^-N$ interaction the high-precision data on the
$\Xi^-$ hypernuclei are needed. It is expected that such data on the missing-mass spectra for the $^{12}$C($K^-$,$K^+$) reaction close to the $\Xi^-$ production threshold and the X-ray data on the level shifts and width broadening of the $\Xi^-$ atomic states in the $\Xi^-$ atoms will become available soon from the E70 and E03 experiments at the J-PARC [32].
It is worthwhile to point out that the situation with the $\Xi^-$ nuclear potential at finite momenta is still
unclear at present, in spite of a lot theoretical activity [22--27] in this field.
In addition to the above-mentioned ($K^-$,$K^+$) reactions on nuclear targets at beam momenta
of 1.6--1.8 GeV/c, the medium modification of the $\Xi^-$ hyperon could be probed directly,
as was shown in Ref. [33], via the another inclusive near-threshold ($K^-$,$\Xi^-$) reactions
on nuclei. This modification can also be sought by studying the cascade hyperon production off nuclei
in photon-induced reactions at energies close to the threshold energy ($\approx$ 2.4 GeV) for
its production off a free target nucleon at rest.
Owing to a negligible strength of initial-state photon interaction in relation to hadron-nucleus interaction,
photon-nucleus reactions appear to be a more useful tool for studying in-medium properties of $\Xi^-$
hyperons and their interactions in cold nuclear matter both at the threshold and at finite momenta.
In this context, the main aim of the present study is to continue the investigation of the in-medium
properties of the $\Xi^-$ hyperons via their photoproduction off nuclei. We give the predictions
for the absolute differential cross sections and their ratios
for near-threshold photoproduction of $\Xi^-$ hyperons off $^{12}$C and $^{184}$W nuclei
at laboratory angles of $\le$ 45$^{\circ}$ by incident photons with energies of 2.5 and 3.0 GeV
within three different scenarios for the effective scalar mean-field nuclear potential that
$\Xi^-$ hyperon feels in the medium. These predictions have been obtained in the framework of the first-collision
model based on a nuclear spectral function and developed in Ref. [33] for the
description of the inclusive $\Xi^-$ hyperon production on nuclei in near-threshold antikaon-induced reactions
and modified to the case of photon-nucleus reactions.
Their comparison with the respective data, which could be collected
in the future dedicated experiment, will provide a deeper insight into the $\Xi^-$
in-medium properties. Information obtained from this comparison will supplement that inferred from
the ($K^-$,$K^+$) reactions on nuclear targets. Such experiment might be performed
at the CEBAF facility, especially in view of the
fact that the exclusive photoproduction of $\Xi^-$ and $\Xi^0$ hyperons in elementary reactions ${\gamma}p \to K^+{K^+}{\Xi^-}$ and ${\gamma}p \to K^+{K^+}{\pi^-}{\Xi^0}$ has already been investigated here for near-threshold photon energies using the CLAS [34--36] and GlueX [37] detectors.

\section*{2 Formalism: direct $\Xi^-$ hyperon production mechanism in nuclei}

\hspace{0.5cm} Since we are interested in the incident photon energy region only up to 3.0 GeV,
we assume that direct production of $\Xi^-$ hyperons in photon-induced reactions in this region
can proceed in the following ${\gamma}p$ and ${\gamma}n$ elementary processes, having
the lowest threshold for their free production ($\approx$ 2.4 GeV):
\begin{equation}
\gamma+p \to K^++K^++\Xi^-,
\end{equation}
\begin{equation}
\gamma+n \to K^++K^0+\Xi^-.
\end{equation}
We can disregard the contribution to the ground state $\Xi^-$=$\Xi(1320)^-$ production from the elementary
channels ${\gamma}N \to KK{\Xi^-}{\pi}$ and ${\gamma}N \to KK{\Xi(1530)^- \to KK{\Xi^-}{\pi^0}}$,
in which the $\Xi^-$ hyperons are produced directly and through the decay of the intermediate first excited state $\Xi(1530)^-$, at considered incident energies owing to their larger thresholds ($\approx$ 2.7 and 2.9 GeV, respectively) in free ${\gamma}N$ collisions.

According [33, 38], we simplify the subsequent calculations via accounting for the in-medium modifications
of the final $K^+$, $K^0$ mesons and $\Xi^-$ hyperons, involved in the production processes (1), (2),
in terms of their average in-medium masses $<m_{K^+}^*>$, $<m_{K^0}^*>$ and $<m_{\Xi^-}^*>$
in the in-medium cross sections of these processes, which are defined as:
\begin{equation}
<m^*_{h}>=m_{h}+U_h\frac{<{\rho_N}>}{{\rho_0}}+V_{{\rm c}h}({\bar r}),
\end{equation}
where $h$ denotes $K^+$, $K^0$ and $\Xi^-$.
Here, $m_{h}$ is the bare hadron mass, $U_h$ is its effective scalar
nuclear potential which it feels at normal nuclear matter
density ${\rho_0}$, $<{\rho_N}>$ is the average nucleon density and $V_{{\rm c}h}({\bar r})$
is the charged hadron Coulomb potential
\footnote{$^)$This potential for the chosen in the present work ratio $<{\rho_N}>/{\rho_0}$
of 0.55 and 0.76 for $^{12}$C and $^{184}$W nuclei of interest for
positively and negatively charged hadrons is about of +3.6 and +17.3 MeV and -3.6 and -17.3 MeV
for these nuclei, respectively.}$^)$
corresponding to the interaction between the point hadron $h$ and the uniformly charged spherical core
of charge $Z$ and taken at a point ${\bar r}$ at which the local nucleon density
is equal to the average density
\footnote{$^)$Since the incident photons interact with the intranuclear nucleons practically
in the whole volume of the nucleus due to the lack of strong initial-state interactions.}$^)$
.
For the $K^+$ scalar potential $U_{K^+}$ we will use the following option: $U_{K^+}=+22$ MeV [39, 40].
The same option will be adopted
for the $K^0$ nuclear potential $U_{K^0}$ in the case of $^{12}$C target nucleus,
i.e.: $U_{K^0}=U_{K^+}=+22$ MeV [33, 40--42]. And in the case of $^{184}$W nucleus for the $K^0$
scalar potential $U_{K^0}$ we will use the following value: $U_{K^0}=+40$ MeV [33, 40, 41].

Now, we specify the effective scalar mean-field $\Xi^-$ hyperon potential $U_{\Xi^-}$ in Eq. (3).
We assume that this potential is the same for both considered target nuclei $^{12}$C and $^{184}$W
\footnote{$^)$It should be pointed out that this assumes a purely isoscalar $\Xi^-$ potential
$U_{\Xi^-}^{\rm isoscalar}(r)$, associated with the total nuclear density $\rho_n(r)+\rho_p(r)$, for both
these nuclei and does not take into account the contribution to the total strong nuclear $\Xi^-$ potential
$U_{\Xi^-}(r)$ from the isovector one $U_{\Xi^-}^{\rm isovector}(r)$, appearing in the neutron-rich nuclei
due to the neutron excess density $\rho_n(r)-\rho_p(r)$. The latter might be nonnegligible in the $^{184}$W
nucleus. As was shown in Ref. [33],
the ratio $U_{\Xi^-}^{\rm isovector}(r)/U_{\Xi^-}^{\rm isoscalar}(r)$ for this nucleus can reach the value of -0.5,
which is nonnegligible.
The above-mentioned means that a more realistic analysis of the $\Xi^-$ hyperon production in heavy
asymmetric targets should take into account in principle not only the isoscalar $\Xi^-$-nucleus potential, but also
and an isovector one. However, since the main difference between the $\Xi^-$ nuclear potentials in light (symmetric)
and heavy (asymmetric) nuclei is coming from the Coulomb forces, which are included in our model, one may hope
that the role played by the real isovector $\Xi^-$ potential in this production will be moderate.
Its rigorous evaluation is beyond the scope of the present study.}$^)$
.
Since the reachable range of the $\Xi^-$
hyperon vacuum momenta at the considered photon energies is about of 0.2--2.5 GeV/c (see below),
one needs to evaluate it also for such in-medium $\Xi^-$ momenta at saturation density $\rho_0$.
In line with the light quark content of the $\Xi^-$ hyperons: $\Xi^-$=$|dss>$, their mean-field scalar
$U_{S{\Xi^-}}$ and vector $U_{V{\Xi^-}}$ potentials are about 1/3 of those
$U_{SN}$ and $U_{VN}$ of a nucleon [43--45] when in-medium nucleon and $\Xi^-$ hyperon velocities
$v^{\prime}_N$ and $v^{\prime}_{\Xi^-}$ with respect to the surrounding nuclear matter are equal to each other,
i.e.,
$$
U_{S{\Xi^-}}(v^{\prime}_{\Xi^-},\rho_N)=\frac{1}{3}U_{SN}(v^{\prime}_{N},\rho_N),
$$
\begin{equation}
U_{V{\Xi^-}}(v^{\prime}_{\Xi^-},\rho_N)=\frac{1}{3}{\alpha}U_{VN}(v^{\prime}_{N},\rho_N);\,\,\,
v^{\prime}_{N}=v^{\prime}_{\Xi^-}.
\end{equation}
The condition that velocities $v^{\prime}_N$ and $v^{\prime}_{\Xi^-}$ are equal to each other
leads to the following relation between the in-medium nucleon $p^{\prime}_N$ and the $\Xi^-$ hyperon $p^{\prime}_{\Xi^-}$ momenta:
\begin{equation}
p^{\prime}_{N}=\frac{<m^*_{N}>}{<m^*_{\Xi^-}>}p^{\prime}_{\Xi^-}.
\end{equation}
But, to simplify the  calculations of the $\Xi^-$--nucleus single-particle potential
(or the so-called Schr${\ddot{\rm o}}$dinger equivalent potential $V_{{\Xi^-}A}^{\rm SEP}$), we will
use in Eq. (5) vacuum nucleon and $\Xi^-$ hyperon masses $m_N$ and $m_{\Xi^-}$ instead of
their average in-medium masses $<m^*_{N}>$ and $<m^*_{\Xi^-}>$.
Then, this potential $V_{{\Xi^-}A}^{\rm SEP}$ is derived from the $\Xi^-$ in-medium energy as [44, 45]:
\begin{equation}
V_{{\Xi^-}A}^{\rm SEP}(p^{\prime}_{\Xi^-},\rho_N)=
\sqrt{\left[m_{\Xi^-}+U_{S{\Xi^-}}(p^{\prime}_{\Xi^-},\rho_N)\right]^2+({p^{\prime}_{\Xi^-}})^2}
+U_{V{\Xi^-}}(p^{\prime}_{\Xi^-},\rho_N)-\sqrt{m_{\Xi^-}^2+({p^{\prime}_{\Xi^-}})^2}.
\end{equation}
The relation between the potentials $U_{\Xi^-}$ and $V_{{\Xi^-}A}^{\rm SEP}$ at density $\rho_0$ is given by
\begin{equation}
U_{\Xi^-}({p^{\prime}_{\Xi^-}})=
\frac{\sqrt{m^2_{\Xi^-}+({p^{\prime}_{\Xi^-}})^2}}{m_{\Xi^-}}V_{{\Xi^-}A}^{\rm SEP}({p^{\prime}_{\Xi^-}}).
\end{equation}
Adopting the following parametrizations for the nucleon potentials $U_{SN}$ and $U_{VN}$
at density $\rho_0$ from [46]
\begin{equation}
U_{SN}(p^{\prime}_{N},\rho_0)=-\frac{494.2272}{1+0.3426\sqrt{p^{\prime}_{N}/p_F}}~{\rm MeV},
\end{equation}
\begin{equation}
U_{VN}(p^{\prime}_{N},\rho_0)=\frac{420.5226}{1+0.4585\sqrt{p^{\prime}_{N}/p_F}}~{\rm MeV}
\end{equation}
(where $p_F=1.35$ fm$^{-1}=$0.2673 GeV/c), we calculated, with accounting for Eqs. (4)--(7)
as well as by multiplying the vector nucleon potential in Eq. (4) by a factor $\alpha$
of $\alpha=$1.068 [42] to get for $\Lambda$ hyperon potential $V_{{\Lambda}A}^{\rm SEP}$ at zero momentum
and at density $\rho_0$ an experimental value of -(32$\pm$2) MeV [47],
the momentum dependences of potentials $V_{{\Xi^-}A}^{\rm SEP}$ and $U_{\Xi^-}$ at saturation density .
The $\Xi^-$ potentials $V_{{\Xi^-}A}^{\rm SEP}$ and $U_{\Xi^-}$, adjusted in this way, are shown in Fig. 1
by dashed and dotted-dotted-dashed curves, respectively. It is seen that these potentials
are attractive for all momenta below of about 0.9 GeV/c with the value of
$V_{{\Xi^-}A}^{\rm SEP}(0)=U_{\Xi^-}(0)\approx$-15 MeV,
whereas they are repulsive for higher momenta and increase monotonically with the growth of $\Xi^-$ momentum.
\begin{figure}[!ht]
\begin{center}
\includegraphics[width=18.0cm]{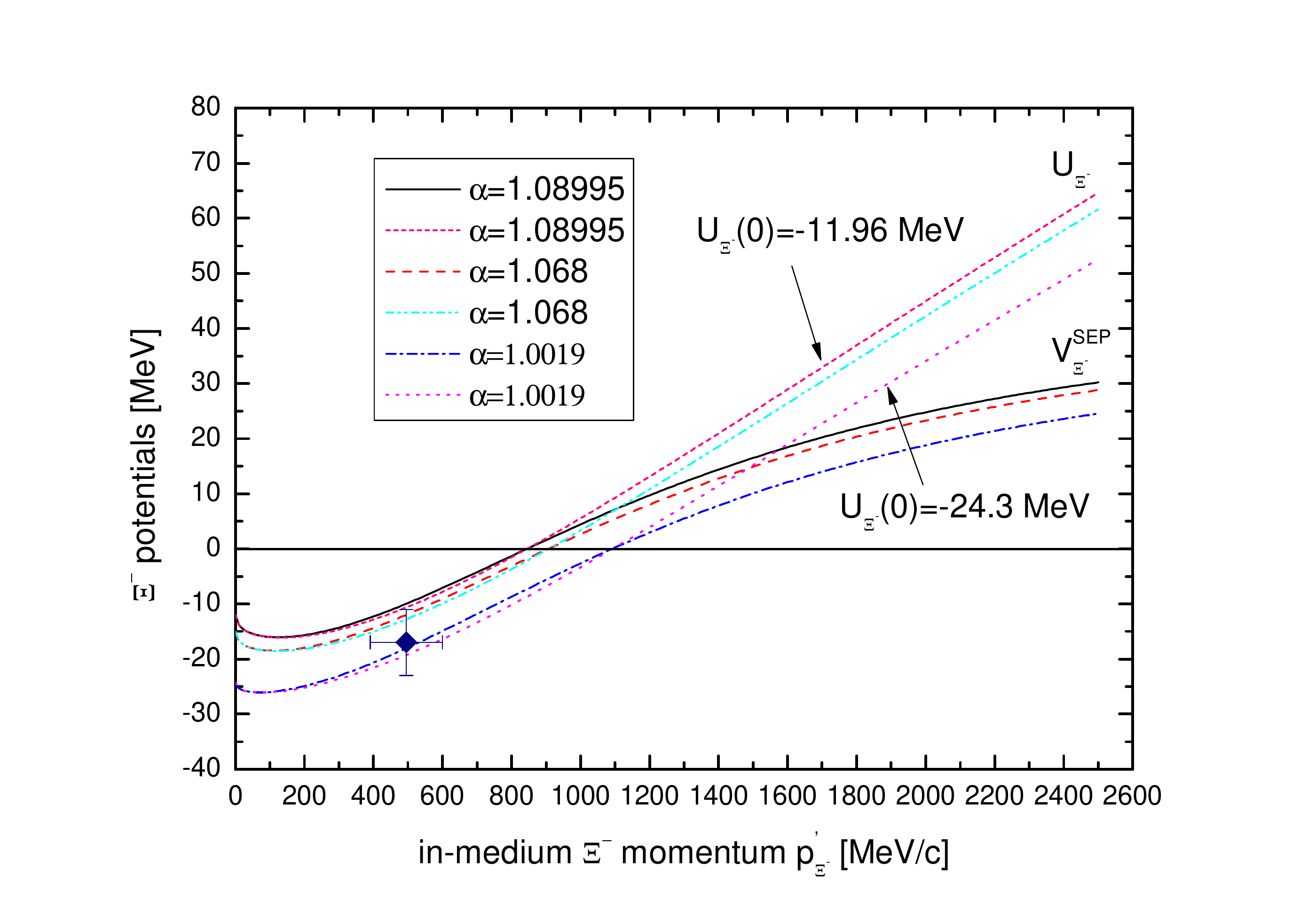}
\vspace*{-2mm} \caption{(Color online) Momentum dependence of the Schr$\ddot{\rm o}$dinger equivalent
and effective scalar ${\Xi^-}$ hyperon potentials at density $\rho_0$. For notation see the text.}
\label{void}
\end{center}
\end{figure}
This value is in good agreement with the $\Xi^-$--nucleus potential depths of about
-16 and -14 MeV, inferred from the analysis of the data on the missing-mass
spectra for the $^{12}$C($K^-$,$K^+$) reaction in the $\Xi^-$-bound state region
collected by the KEK--PS E224 [10] and BNL--AGS E885 [11] Collaborations
at initial momenta of 1.6 and 1.8 GeV/c, correspondingly. Furthermore, it is also well consistent with that of
$\approx$-16 MeV for the in-medium $\Xi^-$ mass shift predicted recently in Ref. [48] in the framework
of the chiral soliton model. In addition, results for two $\Xi^-$ potentials
$V_{{\Xi^-}A}^{\rm SEP}$, $U_{\Xi^-}$, obtained by multiplying the vector nucleon potential in Eq. (4)
by factors $\alpha$ of $\alpha=$1.08995 and $\alpha=$1.0019 to get for the $\Xi^-$ hyperon
single-particle potential $V_{{\Xi^-}A}^{\rm SEP}$ at zero momentum and at density $\rho_0$ the values of -11.96
and -24.3 MeV, are given in Fig. 1 as well. These values were
deduced very recently in Refs. [8] and [9], correspondingly, from the analysis of the $\Xi^-$
capture events in light nuclear emulsion identified in KEK and J-PARC experiments [5--7] within the
quark mean-field model and in the framework of a $t{\rho}$ optical potential method. These potentials are depicted
in Fig. 1 by solid, short-dashed and dotted-dashed, dotted lines, respectively.
It is seen that now the former potentials are attractive at all momenta below of about 0.85 GeV/c, whereas they
are repulsive at higher momenta. Contrary to this case, the latter potentials turn to repulsion at fairly high
momenta, around 1.1 GeV/c. At momenta below $\approx$ 0.8 GeV/c these potentials, respectively, are more weakly
and are more strongly attractive than those, calculated for a correction factor $\alpha=$1.068. And they,
correspondingly, are more strongly and are more weakly repulsive than the latters at $\Xi^-$ momenta above
$\approx$ 1.1 GeV/c. Since currently there are essential
uncertainties in determining the $\Xi^-$ nuclear potential both at threshold and at finite momenta below 1 GeV/c
(cf. [23--31]), it is natural to adopt for certainty for the quantity $U_{\Xi^-}$ in the present work two momentum
dependences denoted in Fig. 1 by the short-dashed and dotted lines, respectively, and corresponding at zero
momentum relative to the bulk matter to the values of $U_{\Xi^-}(0)=-11.96$ MeV and $U_{\Xi^-}(0)=-24.3$ MeV.
This will allow us to explore the sensitivity of the $\Xi^-$ production cross sections from the direct processes
(1), (2) to the potential $U_{\Xi^-}$. Moreover, to extend the range of applicability of our model, we will also
ignore it in our calculations. Thus, the results will be given for three scenarios
for a $\Xi^-$ potential: i) momentum-dependent potential with a value of $U_{\Xi^-}(0)=-24.3$ MeV,
ii) momentum-dependent potential with a value of $U_{\Xi^-}(0)=-11.96$ MeV
and iii) zero potential at all $\Xi^-$ momenta, achievable in the $\Xi^-$ production in
${\gamma}A$ interactions at photon energies of interest. It should be pointed out that nowadays there is no
experimental information about the momentum dependence of the $\Xi^-$ nuclear potential at finite momenta.
To some extent, such "experimental" information on the possible strength
of the $\Xi^-$ nuclear potential for $\Xi^-$--$^8$Li system for finite momenta follows from the
analysis [17] of the BNL--E906 measurement [14] of the spectrum of the $^9$Be($K^-$,$K^+$) reaction at initial $K^-$
momentum of 1.8 GeV/c in the $\Xi^-$ quasi-free region within the DWIA method.
The strength of (-17$\pm$6) MeV for this potential, taken in the Woods-Saxon form, was extracted from this analysis
in the momentum transfer region $q$ $\simeq$ 390--600 MeV/c [17]. If we suppose that $p^{\prime}_{\Xi^-}=q$, then this
value, shown as data point in Fig. 1, can be considered as the central depth of the $\Xi^-$--nucleus potential
in the momentum range of $\simeq$ 390--600 MeV/c and, therefore, it can be compared with the aforementioned
predictions given in this figure. One can see that only calculation, which corresponds to $V_{{\Xi^-}A}^{\rm SEP}(0)=-24.3$ MeV, is fairly consistent with the result of [17] over the momentum range of 390--600 MeV/c.
More detailed information on the $\Xi^-$
potential at finite momenta could be in particular obtained from comparison the results of the present model calculations with the respective precise experimental data on direct $\Xi^-$ photoproduction on nuclei.
These data could be collected in future dedicated experiment at the CEBAF facility using high-intensity photon beams.

      The total energy $E^\prime_{h}$ of the hadron in nuclear medium is linked with
its average effective mass $<m^*_{h}>$ and its in-medium momentum
${\bf p}^{\prime}_{h}$ as follows [33, 38]:
\begin{equation}
E^\prime_{h}=\sqrt{({\bf p}^{\prime}_{h})^2+(<m^*_{h}>)^2}.
\end{equation}
The hadron in-medium momentum ${\bf p}^{\prime}_{h}$ is related to its vacuum momentum ${\bf p}_{h}$
in the following way [33, 38]:
\begin{equation}
E^\prime_{h}=\sqrt{({\bf p}^{\prime}_{h})^2+(<m^*_{h}>)^2}=
\sqrt{{\bf p}^2_{h}+m^2_{h}}=E_{h},
\end{equation}
where $E_{h}$ is the hadron total energy in vacuum. For the considered above $\Xi^-$ hyperon momentum-dependent potentials its in-medium momentum $p^{\prime}_{\Xi^-}$, with accounting for the relation (3), is the root of an
equation (11) at given vacuum momentum $p_{\Xi^-}$. This root can be found by using the respective numerical
procedure. But, to simplify the calculation of the $\Xi^-$ production in ${\gamma}A$ interactions, we will
determine these potentials at $\Xi^-$ hyperon vacuum momentum $p_{\Xi^-}$ and after that, employing Eq. (11),
its in-medium momentum $p^{\prime}_{\Xi^-}$. For our aims, as we checked, this is well justified.

The $\Xi^-$ hyperon--nucleon elastic cross section $\sigma^{\rm el}_{\Xi^-N}$ is expected to be $\sim$ 10 mb in the $\Xi^-$ momentum range of 0.2--2.5 GeV/c [28, 29, 31, 49--53] studied in the present work.
Its mean "free" path $\lambda^{\rm el}_{\Xi^-}$ up to elastic scattering inside the nucleus, defined as
$\lambda^{\rm el}_{\Xi^-}=1/(<\rho_N>\sigma^{\rm el}_{\Xi^-N})$, for this cross section and for average
nucleon densities of $<\rho_N>=0.55\rho_0$ ($^{12}$C) and $<\rho_N>=0.76\rho_0$ ($^{184}$W), $\rho_0=$0.16 fm$^{-3}$
amounts $\approx$ 11 fm for $^{12}$C and $\approx$ 8 fm for $^{184}$W.
The radii of $^{12}$C and $^{184}$W, which are approximately 3 and 7.4 fm, are, respectively, less than these values.
Therefore, we can ignore elastic $\Xi^-N$ scatterings in the present study.
Since in-medium threshold energies $\sqrt{s^*_{\rm th}}=2<m_{K^+}^*>+<m_{\Xi^-}^*>$ and
$\sqrt{{\tilde s}^*_{\rm th}}=<m_{K^+}^*>+<m_{K^0}^*>+<m_{\Xi^-}^*>$
of the elementary reactions (1) and (2) are practically equal to each other for both considered target nuclei and
the free total cross sections of these reactions are also practically equal to each other [54]
at beam energies of interest, we assume that their in-medium total cross sections are practically equal to each other
as well. Then, neglecting the attenuation of the initial photon beam in the nuclear matter
and describing the $\Xi^-$ hyperon final-state absorption by
the in-medium total inelastic $\Xi^-N$ cross section $\sigma_{{\Xi^-}N}^{\rm in}(p^{\prime}_{\Xi^-})$,
we express the inclusive differential cross section for direct production of $\Xi^-$ hyperons
off nuclei in the processes (1), (2) as [55]:
\begin{equation}
\frac{d\sigma_{{\gamma}A \to {\Xi^-}X}^{({\rm dir})}
(E_{\gamma},{\bf p}_{\Xi^-})}
{d{\bf p}_{\Xi^-}}=I_{V}[A]
\left<\frac{d\sigma_{{\gamma}p\to K^+K^+{{\Xi^-}}}({\bf p}_{\gamma},
{\bf p}^{\prime}_{{\Xi^-}})}{d{\bf p}^{\prime}_{{\Xi^-}}}\right>_A\frac{d{\bf p}^{\prime}_{{\Xi^-}}}
{d{\bf p}_{{\Xi^-}}},
\end{equation}
where
\begin{equation}
I_{V}[A]=2{\pi}A\int\limits_{0}^{R}r_{\bot}dr_{\bot}
\int\limits_{-\sqrt{R^2-r_{\bot}^2}}^{\sqrt{R^2-r_{\bot}^2}}dz
\rho(\sqrt{r_{\bot}^2+z^2})\exp{\left[-\sigma_{{\Xi^-}N}^{\rm in}(p^{\prime}_{\Xi^-})A
\int\limits_{z}^{\sqrt{R^2-r_{\bot}^2}}
\rho(\sqrt{r_{\bot}^2+x^2})dx\right]};
\end{equation}
\begin{equation}
\sigma_{{\Xi^-}N}^{\rm in}(p^{\prime}_{\Xi^-})=
\frac{Z}{A}\sigma_{{\Xi^-}p}^{\rm in}(p^{\prime}_{\Xi^-})+
\frac{N}{A}\sigma_{{\Xi^-}n}^{\rm in}(p^{\prime}_{\Xi^-})
\end{equation}
and
\begin{equation}
\left<\frac{d\sigma_{{\gamma}p\to K^+K^+{\Xi^-}}({\bf p}_{\gamma},{\bf p}^{\prime}_{\Xi^-})}
{d{\bf p}^{\prime}_{\Xi^-}}\right>_A=
\int\int
P_A({\bf p}_t,E)d{\bf p}_tdE
\end{equation}
$$
\times
\left\{\frac{d\sigma_{{\gamma}p\to K^+K^+{\Xi^-}}[\sqrt{s^*},<m_{K^+}^*>,<m_{K^+}^*>,
<m_{\Xi^-}^*>,{\bf p}^{\prime}_{\Xi^-}]}
{d{\bf p}^{\prime}_{\Xi^-}}\right\},
$$
\begin{equation}
  s^*=(E_{\gamma}+E_t)^2-({\bf p}_{\gamma}+{\bf p}_t)^2,
\end{equation}
\begin{equation}
   E_t=M_A-\sqrt{(-{\bf p}_t)^2+(M_{A}-m_{N}+E)^{2}}.
\end{equation}
Here,
$d\sigma_{{\gamma}p\to K^+{K^+}{\Xi^-}}[\sqrt{s^*},<m_{K^+}^*>,<m_{K^+}^*>,<m_{\Xi^-}^*>,{\bf p}^{\prime}_{\Xi^-}]
/d{\bf p}^{\prime}_{\Xi^-}$
is the in-medium differential cross section for the production of
$\Xi^-$ hyperon with modified mass $<m_{\Xi^-}^*>$ and in-medium momentum
${\bf p}^{\prime}_{{\Xi^-}}$ in process (1) at the ${\gamma}p$ center-of-mass energy $\sqrt{s^*}$.
The $K^+$ mesons are produced in this process with the medium modified mass $<m_{K^+}^*>$ as well.
$E_{\gamma}$ and ${\bf p}_{\gamma}$ are the total energy and momentum of the incident photon;
$\rho({\bf r})$ and $P_A({\bf p}_t,E)$ are the local nucleon density and the
spectral function of the target nucleus A normalized to unity
(the information about these quantities can be found in Refs. [39, 56--58]);
${\bf p}_t$ and $E$ are the momentum and removal energy of the struck target proton
participating in the production process (1);
$Z$ and $N$ are the numbers of protons and neutrons in the target nucleus ($A=Z+N$),
$M_{A}$  and $R$ are its mass and radius; $m_N$ is the bare nucleon mass.
In Eq. (12) we assume that the $\Xi^-$
production cross sections in ${\gamma}p$ and ${\gamma}n$ reactions (1) and (2) are the same [54] as well as
that the way of the produced $\Xi^-$ hyperon from the production point
in the nucleus to a free space is not perturbed
by a relatively weak nuclear fields being considered in the present work, by the
attractive Coulomb potential and by rare ${\Xi^-}N$ elastic scatterings. Therefore,
the in-medium hyperon momentum ${\bf p}^{\prime}_{\Xi^-}$
is assumed to be parallel to the vacuum one ${\bf p}_{\Xi^-}$
and they are related by the Eq. (11).

   According to [33, 38], we assume that the differential cross section
$d\sigma_{{\gamma}p \to K^+{K^+}{\Xi^-}}[\sqrt{s^*},<m_{K^+}^*>,<m_{K^+}^*>,<m_{\Xi^-}^*>,{\bf p}^{\prime}_{\Xi^-}]
/d{\bf p}^{\prime}_{\Xi^-}$ for $\Xi^-$ hyperon production via reaction (1) in a nuclear medium is equivalent to
the respective free space cross section calculated for the off-shell kinematics of this reaction
and for the final $K^+$ mesons and $\Xi^-$ hyperon in-medium masses $<m_{K^+}^*>$
and $<m_{\Xi^-}^*>$. In our calculations,
this differential cross section has been described according to the three-body phase space
at beam energies of interest (cf. [35]
\footnote{$^)$ Presented here experimental angular distributions of the $\Xi^-$ and $K^+$
in the photon-proton center-of-mass frame show practically an isotropic behavior at photon energies below
3.0 GeV.}$^)$
):
\begin{equation}
\frac{d\sigma_{{\gamma}p \to K^+{K^+}{\Xi^-}}[\sqrt{s^*},<m_{K^+}^*>,<m_{K^+}^*>,<m_{\Xi^-}^*>,{\bf p}^{\prime}_{\Xi^-}]}
{d{\bf p}^{\prime}_{\Xi^-}}=
\frac{\pi}{4}
\end{equation}
$$
\times
\frac{\sigma_{{\gamma}p \to K^+{K^+}{\Xi^-}}(E^*_{\gamma},E^{*{\rm th}}_{\gamma})}
{I_3[s^*,<m_{K^+}^*>,<m_{K^+}^*>,<m_{\Xi^-}^*>]E^{\prime}_{\Xi^-}}
\frac{\lambda[s^*_{K^+K^+},(<m_{K^+}^*>)^{2},(<m_{K^+}^*>)^{2}]}{s^*_{K^+K^+}},
$$
where
\begin{equation}
I_3[s^*,a,b,c]=\left(\frac{\pi}{2}\right)^2\int\limits_{(b+c)^2}^{(\sqrt{s^*}-a)^2}
\frac{\lambda[x,b^{2},c^{2}]}{x}\frac{\lambda[s^*,x,a^{2}]}{s^*}dx,
\end{equation}
\begin{equation}
\lambda(x,y,z)=\sqrt{{\left[x-({\sqrt{y}}+{\sqrt{z}})^2\right]}{\left[x-
({\sqrt{y}}-{\sqrt{z}})^2\right]}},
\end{equation}
\begin{equation}
s^*_{K^+K^+}=s^*+(<m_{\Xi^-}^*>)^2-2(E_{\gamma}+E_t)E^{\prime}_{\Xi^-}+2({\bf p}_{\gamma}+{\bf p}_t)
{\bf p}^{\prime}_{\Xi^-},
\end{equation}
and
\begin{equation}
E^*_{\gamma}=\frac{s^*-m_p^2}{2m_p},\,\,\,\,\,E^{*{\rm th}}_{\gamma}=\frac{s^*_{\rm th}-m_p^2}{2m_p}.
\end{equation}
Here, $\sigma_{{\gamma}p \to K^+{K^+}{\Xi^-}}(E^*_{\gamma},E^{*{\rm th}}_{\gamma})$ is the
in-medium total cross section of process (1). In line with the aforesaid, it is equivalent to the free cross section
$\sigma_{{\gamma}p \to K^+{K^+}{\Xi^-}}(E_{\gamma},E^{{\rm th}}_{\gamma})$, in which the vacuum incident
$E_{\gamma}$ and threshold $E^{{\rm th}}_{\gamma}$  energies
\footnote{$^)$ Let us remind that $E^{{\rm th}}_{\gamma}=$2.372 GeV.}$^)$
are replaced by the in-medium ones $E^*_{\gamma}$ and $E^{*{\rm th}}_{\gamma}$, defined by the Eq. (22).
\begin{figure}[htb]
\begin{center}
\includegraphics[width=18.0cm]{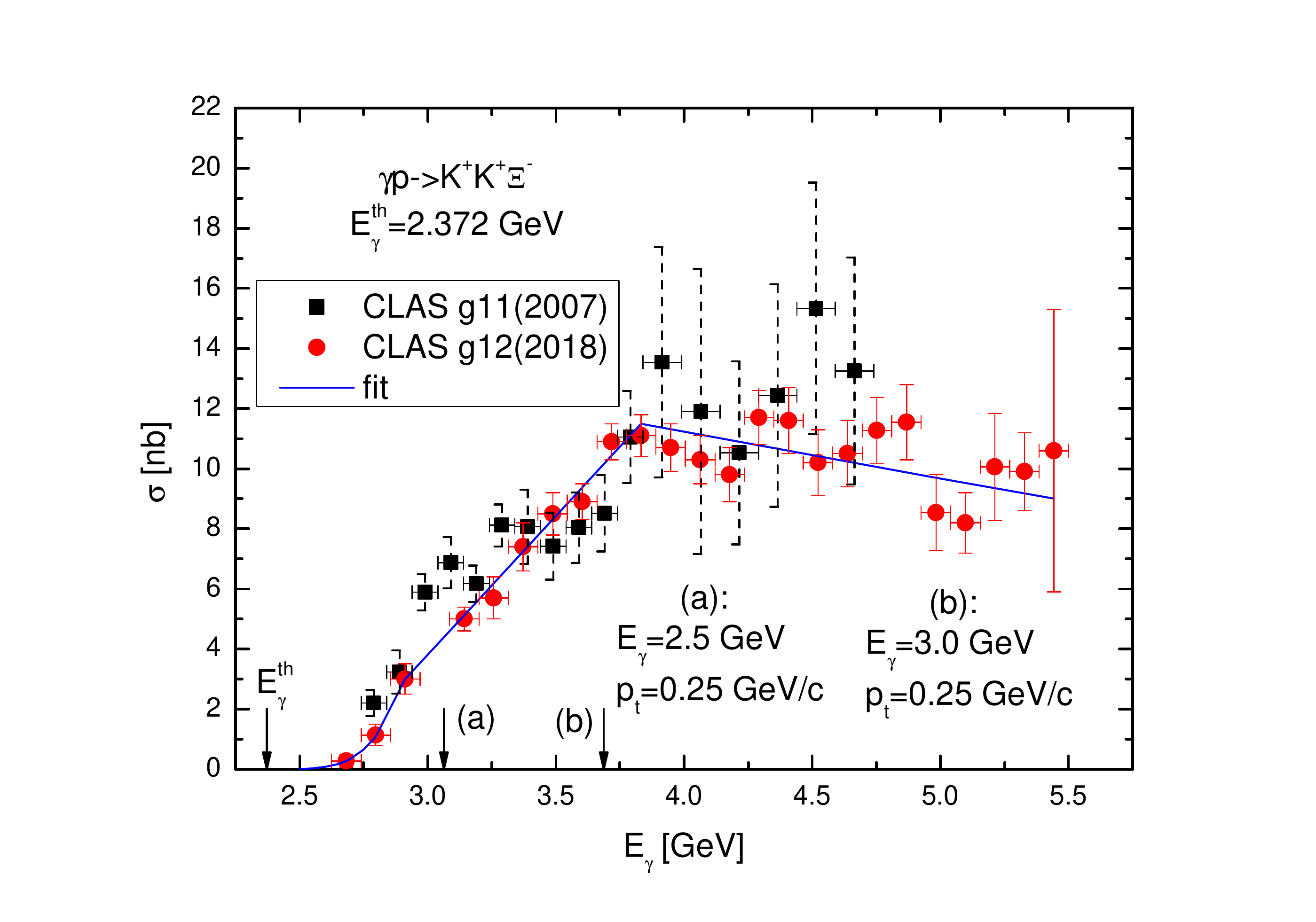}
\vspace*{-2mm} \caption{(Color online) Total cross section for the reaction ${\gamma}p \to {K^+}{K^+}{\Xi^-}$
as function of the incident photon energy $E_{\gamma}$. The left arrow indicates the free threshold energy for
this reaction. The middle and right arrows indicate the effective photon energies $E^*_{\gamma}=3.063$ GeV and
$E^*_{\gamma}=3.686$ GeV corresponding to the incident photon energies of 2.5 and 3.0 GeV, respectively,
and a target proton bound in $^{12}$C by 16 MeV and with momentum of 250 MeV/c.
The latter one is directed opposite to the incoming photon beam. For the rest of the notation see the text.}
\label{void}
\end{center}
\end{figure}

     The currently available experimental information on the free cross section
$\sigma_{{\gamma}p \to K^+{K^+}{\Xi^-}}(E_{\gamma},E^{{\rm th}}_{\gamma})$ from CLAS g11 [35] and
CLAS g12 [36] experiments, performed at Jefferson Lab in the photon low-energy range $E_{\gamma} \le 5.44$ GeV
\footnote{$^)$ The author thanks I. Strakovsky for sending this information from CLAS g11 experiment to him.}$^)$,
can be fitted as:
\begin{equation}
\sigma_{{\gamma}p \to K^+{K^+}{\Xi^-}}(E_{\gamma},E^{{\rm th}}_{\gamma})
\end{equation}
$$
=\left\{
\begin{array}{llll}
	41.733\left[(E_{\gamma}-E^{{\rm th}}_{\gamma})/{\rm GeV}\right]^{4.278}~[{\rm nb}]
	&\mbox{for $E^{{\rm th}}_{\gamma} \le E_{\gamma} \le 2.9125~{\rm GeV}$}, \\
	&\\
        -23.909+9.239(E_{\gamma}/{\rm GeV})~[{\rm nb}]
	&\mbox{for $2.9125~{\rm GeV} < E_{\gamma} \le 3.8325~{\rm GeV}$}, \\
    &\\
     17.452-1.553(E_{\gamma}/{\rm GeV})~[{\rm nb}]
    &\mbox{for $3.8325~{\rm GeV} < E_{\gamma} \le 5.4425~{\rm GeV}$}.
\end{array}
\right.	
$$
It is seen from Fig. 2 that the parametrization (23) (solid line) fits quite well the existing data
[35, 36] (solid squares and circles).
One can also see that the off-shell cross section $\sigma_{{\gamma}p \to K^+{K^+}{\Xi^-}}$, calculated
in line with Eqs. (16), (17), (22) and (23) for photon energies of 2.5 and 3.0 GeV, a target proton bound in
$^{12}$C by 16 MeV, and with its internal momentum of 250 MeV/c directed opposite to the initial photon beam,
is about 5 and 10 nb, respectively. This opens up the possibility of measuring the sizable $\Xi^-$
hyperon production cross sections in $(\gamma,\Xi^-)$ reactions near threshold at the CEBAF facility.
It is worth noting that preliminary results for the total cross section of the reaction
${\gamma}p \to K^+{K^+}{\Xi^-}$ at higher photon energies have been recently obtained by the GlueX
Collaboration at the upgraded up to 12 GeV CEBAF facility [37].

At the considered initial photon energies $\Xi^-$ hyperons are mainly produced in ${\gamma}N$ interactions
at small angles in the laboratory frame
\footnote{$^)$ For instance, the maximum angle of $\Xi^-$ hyperon production off a free proton at rest
in reaction (1) is about 20.3$^{\circ}$ at the photon energy of 3.0 GeV.}$^)$
.
Therefore, we will calculate the $\Xi^-$ momentum
differential distributions from $^{12}$C and $^{184}$W target nuclei
for the laboratory solid angle
${\Delta}{\bf \Omega}_{\Xi^-}$=$0^{\circ} \le \theta_{\Xi^-} \le 45^{\circ}$,
and $0 \le \varphi_{\Xi^-} \le 2{\pi}$.
Here, $\theta_{\Xi^-}$ is the polar angle of vacuum momentum ${\bf p}_{\Xi^-}$ in the laboratory system
with z-axis directed along the momentum ${\bf p}_{\gamma}$ of the incident photon beam and
$\varphi_{\Xi^-}$ is the azimuthal angle of the $\Xi^-$ momentum ${\bf p}_{\Xi^-}$
in this system. We will require - as this has been done in Ref. [59] in the analysis of the data
on $\eta^{\prime}$ meson photoproduction on $^{12}$C target nucleus - that the produced with vacuum momentum
${\bf p}_{\Xi^-}$ $\Xi^-$ hyperon goes backward in the c.m.s. of the incident photon beam and
a target nucleon at rest to guaranty  that it will have relatively low momentum in the laboratory system,
at which the sensitivity of the $\Xi^-$ hyperon yield to its in-medium modification is expected to be enhanced,
Then, accounting for Eq. (12), we can express the respective differential cross section for direct photon-induced $\Xi^-$ hyperon production off nuclei from the processes (1), (2),
corresponding to this solid angle and to this kinematical condition, in the form:
\begin{equation}
\frac{d\sigma_{{\gamma}A\to {\Xi^-}X}^{({\rm dir})}
(E_{\gamma},p_{\Xi^-})}{dp_{\Xi^-}}=
\int\limits_{{\Delta}{\bf \Omega}_{\Xi^-}}^{}d{\bf \Omega}_{\Xi^-}
\frac{d\sigma_{{\gamma}A\to {\Xi^-}X}^{({\rm dir})}
(E_{\gamma},{\bf p}_{\Xi^-})}{d{\bf p}_{\Xi^-}}p_{\Xi^-}^2Q(\cos{\theta^{\rm cm}_{\Xi^-}})
\end{equation}
$$
=2{\pi}\left(\frac{p_{\Xi^-}}{p^{\prime}_{\Xi^-}}\right)I_{V}[A]
\int\limits_{\cos45^{\circ}}^{1}d\cos{{\theta_{\Xi^-}}}
\left<\frac{d\sigma_{{\gamma}p\to K^+{K^+}{\Xi^-}}(p_{\gamma},p^{\prime}_{\Xi^-},\theta_{\Xi^-})}
{dp^{\prime}_{\Xi^-}d{\bf \Omega}_{\Xi^-}}\right>_AQ(\cos{\theta^{\rm cm}_{\Xi^-}}),
$$
where the "$\Xi^-$ c.m.s. forward emission eliminating" factor $Q(\cos{\theta^{\rm cm}_{\Xi^-}})$ is
defined in the following way:
\begin{equation}
Q(\cos{\theta^{\rm cm}_{\Xi^-}})
=\left\{
\begin{array}{lll}
	 1
	&\mbox{for $-1 \le \cos{\theta^{\rm cm}_{\Xi^-}} < 0$}, \\
	&\\
        0
	&\mbox{{\rm otherwise}}
\end{array}
\right.	
\end{equation}
and
\begin{equation}
\cos{\theta^{\rm cm}_{\Xi^-}}=\frac{p_{\gamma}p_{\Xi^-}\cos{\theta_{\Xi^-}}-E_{\gamma}E_{\Xi^-}+
E^{\rm cm}_{\gamma}E^{\rm cm}_{\Xi^-}}{p^{\rm cm}_{\gamma}p^{\rm cm}_{\Xi^-}}.
\end{equation}
Here,
\begin{equation}
E^{\rm cm}_{\gamma}=p^{\rm cm}_{\gamma}=\frac{1}{2\sqrt{s}}\lambda(s,0,m^2_p),
\end{equation}
\begin{equation}
E^{\rm cm}_{\Xi^-}=\gamma_{\rm cm}(E_{\Xi^-}-p_{\Xi^-}v_{\rm cm}\cos{\theta_{\Xi^-}}),\,\,\,
p^{\rm cm}_{\Xi^-}=\sqrt{(E^{\rm cm}_{\Xi^-})^2-m^2_{\Xi^-}},
\end{equation}
\begin{equation}
\gamma_{\rm cm}=(E_{\gamma}+m_p)/\sqrt{s},\,\,\,v_{\rm cm}=p_{\gamma}/(E_{\gamma}+m_p),\,\,\,
s=(E_{\gamma}+m_p)^2-p^2_{\gamma}.
\end{equation}
In the following, we will present mainly the calculations based on the Eq. (24).
For completeness and for seeing the role of introduced $\Xi^-$ phase space limitations, we will also give
below in Figs. 5, 6, 9, 10 the results of calculations, obtained assuming in this equation the
"phase space eliminating" factor $Q$ to equal to 1. They will be labeled here as "no limitations" case.
We will also consider the impact of the $\Xi^-$ modification in nuclear
matter on the momentum dependence of such relative observable -- the transparency ratio $T_A$
defined as the ratio between the inclusive $\Xi^-$ differential production cross section (24) on a heavy
nucleus and a light one ($^{12}$C):
\begin{equation}
T_A(E_{\gamma},p_{\Xi^-})=\frac{12}{A}\frac{d\sigma_{{\gamma}A\to {\Xi^-}X}^{({\rm dir})}
(E_{\gamma},p_{\Xi^-})/dp_{\Xi^-}}
{d\sigma_{{\gamma}{\rm C}\to {\Xi^-}X}^{({\rm dir})}
(E_{\gamma},p_{\Xi^-})/dp_{\Xi^-}}.
\end{equation}
For the total free inelastic $\Xi^-p$ and $\Xi^-n$ cross sections
$\sigma_{{\Xi^-}p}^{\rm in}$ and $\sigma_{{\Xi^-}n}^{\rm in}$, which enter into Eq. (14) and which will be
used in our calculations of $\Xi^-$ production in ${\gamma}A$ reactions, in Ref. [33] the
following laboratory $\Xi^-$ momentum $p_{\Xi^-}$ dependences have been suggested at $\Xi^-$
momenta below 0.8 GeV/c
\begin{equation}
\sigma_{{\Xi^-}p}^{\rm in}(p_{\Xi^-})=\left\{
\begin{array}{ll}
	2.417/(p_{\Xi^-})^{1.8106}~[{\rm mb}]
	&\mbox{for $0.15 \le p_{\Xi^-} \le 0.4~{\rm GeV/c}$}, \\
	&\\
                   12.7~[{\rm mb}]
	&\mbox{for $0.4 < p_{\Xi^-} \le 0.8~{\rm GeV/c}$},
\end{array}
\right.	
\end{equation}
where momentum $p_{\Xi^-}$ is expressed in GeV/c. And
\begin{equation}
\sigma_{{\Xi^-}n}^{\rm in}(p_{\Xi^-})=23.448\left(\sqrt{s_{\Xi^-}(p_{\Xi^-})}-\sqrt{s_0}\right)^{0.353}~[{\rm mb}].
\end{equation}
Here,
\begin{equation}
s_{\Xi^-}(p_{\Xi^-})=(E_{\Xi^-}+m_N)^2-p_{\Xi^-}^2
\end{equation}
and $\sqrt{s_0}$=$m_{\Lambda}$+$m_{\Sigma^-}$=2.313132 GeV is the free threshold energy for the
$\Xi^-n \to {\Lambda}\Sigma^-$ reaction. For the in-medium $\Xi^-p$ and $\Xi^-n$ inelastic cross sections
we adopt Eqs. (31) and (32), in which one needs to make only the substitution $p_{\Xi^-} \to p^{\prime}_{\Xi^-}$.
We extend the application of the above dependences to higher $\Xi^-$ momenta of present interest since they
fulfill available scarce experimental constraints [60] on the ${\Xi^-}N$ inelastic cross section at these momenta
(cf. [61]).
\begin{figure}[htb]
\begin{center}
\includegraphics[width=18.0cm]{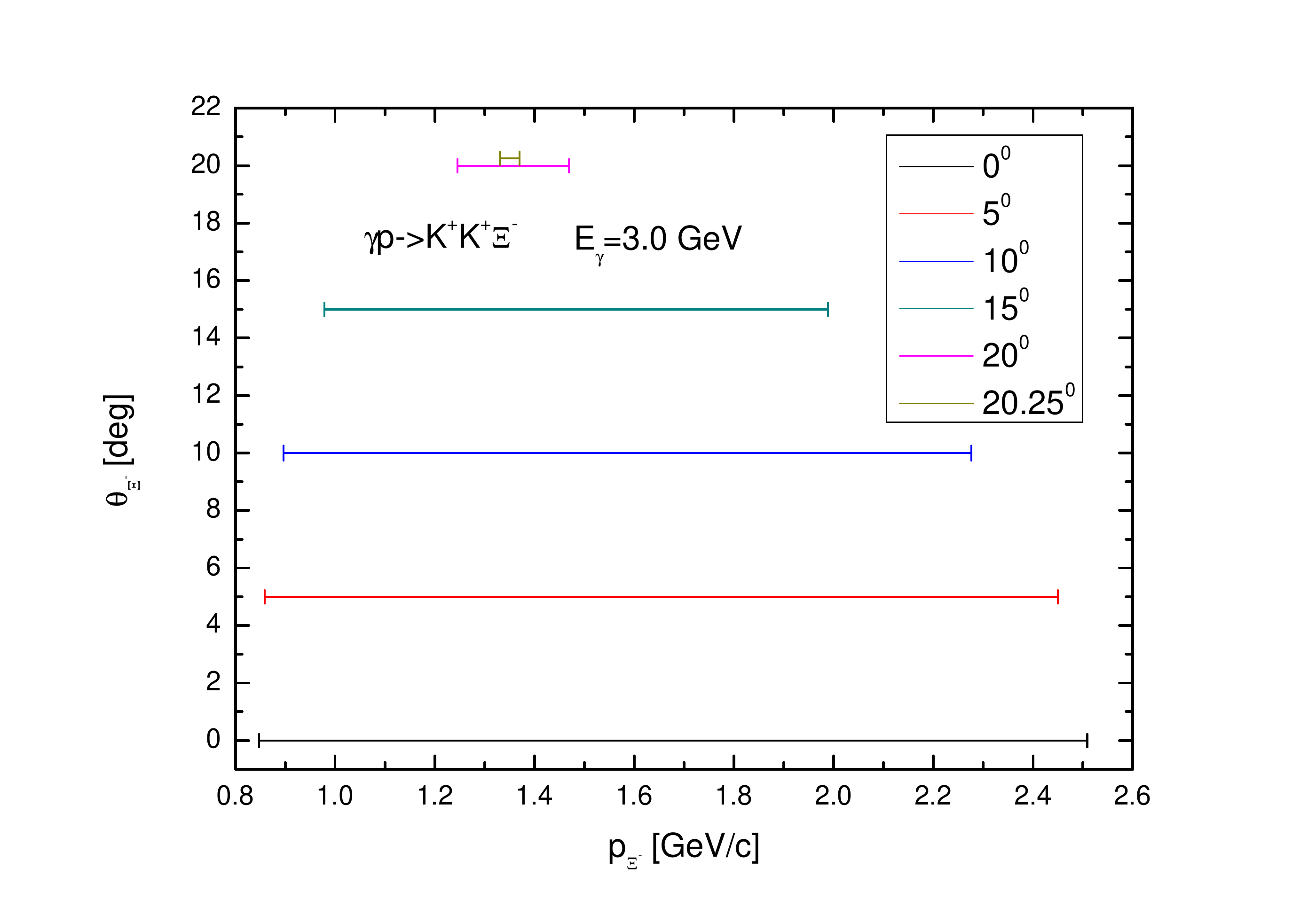}
\vspace*{-2mm} \caption{(Color online) The kinematically allowed momenta of $\Xi^-$ hyperons in the laboratory frame
in the reaction ${\gamma}p \to {K^+}{K^+}{\Xi^-}$ at polar angles, indicated in the inset, for an incident photon energy of 3.0 GeV. The target proton is free and at rest.}
\label{void}
\end{center}
\end{figure}
\begin{figure}[htb]
\begin{center}
\includegraphics[width=18.0cm]{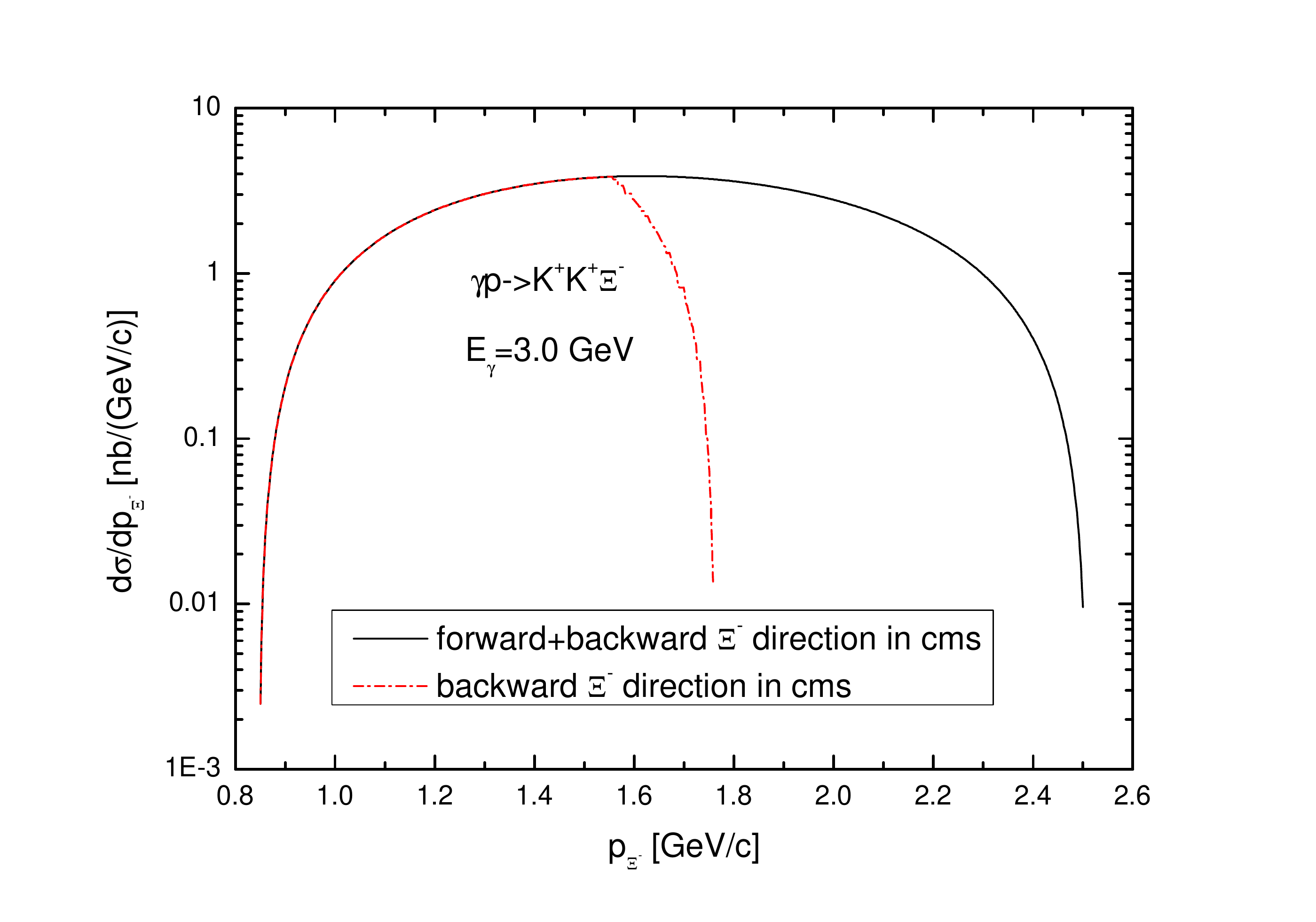}
\vspace*{-2mm} \caption{(Color online) $\Xi^-$ momentum spectrum of the reaction ${\gamma}p \to {K^+}{K^+}{\Xi^-}$
in the full available laboratory polar angular range at an incident photon energy of 3.0 GeV, corresponding to the emission of $\Xi^-$ in the forward+backward (solid curve) and backward (dotted-dashed curve) directions in the center-of-mass system of the incident photon beam and a free target proton at rest.}
\label{void}
\end{center}
\end{figure}
It is interesting to note that, as is easy to see, the integral (13) for the quantity $I_V[A]$ can be
transformed to a simpler expression:
\begin{equation}
I_{V}[A]=\frac{\pi}{\sigma_{{\Xi^-}N}^{\rm in}(p^{\prime}_{\Xi^-})}\int\limits_{0}^{R^2}dr_{\bot}^2
\left(1-e^{-A\sigma_{{\Xi^-}N}^{\rm in}(p^{\prime}_{\Xi^-})\int\limits_{-\sqrt{R^2-r_{\bot}^2}}^{\sqrt{R^2-r_{\bot}^2}}
\rho(\sqrt{r_{\bot}^2+x^2})dx}\right),
\end{equation}
which in the cases of Gaussian nuclear density ($\rho({\bf r})=(b/\pi)^{3/2}\exp{(-br^2)}$)
and a uniform nucleon density for a
nucleus of a radius $R=r_0A^{1/3}$ with a sharp boundary is reduced to even more simple forms:
\begin{equation}
I_V[A]=\frac{A}{x_G}\int\limits_{0}^{1}\frac{dt}{t}\left(1-e^{-x_Gt}\right), \,\,\,\,
x_G=A\sigma_{{\Xi^-}N}^{\rm in}(p^{\prime}_{\Xi^-})b/\pi
\end{equation}
and
\begin{equation}
I_V[A]=\frac{3A}{2a_1}\left\{1-\frac{2}{a_1^2}[1-(1+a_1)e^{-a_1}]\right\}, \,\,\,\,
a_1=3A\sigma_{{\Xi^-}N}^{\rm in}(p^{\prime}_{\Xi^-})/2{\pi}R^2,
\end{equation}
respectively.
The simple formulas (35) and (36) allow one to easily estimate the differential cross section (12)
at above threshold energies.
\begin{figure}[!h]
\begin{center}
\includegraphics[width=18.0cm]{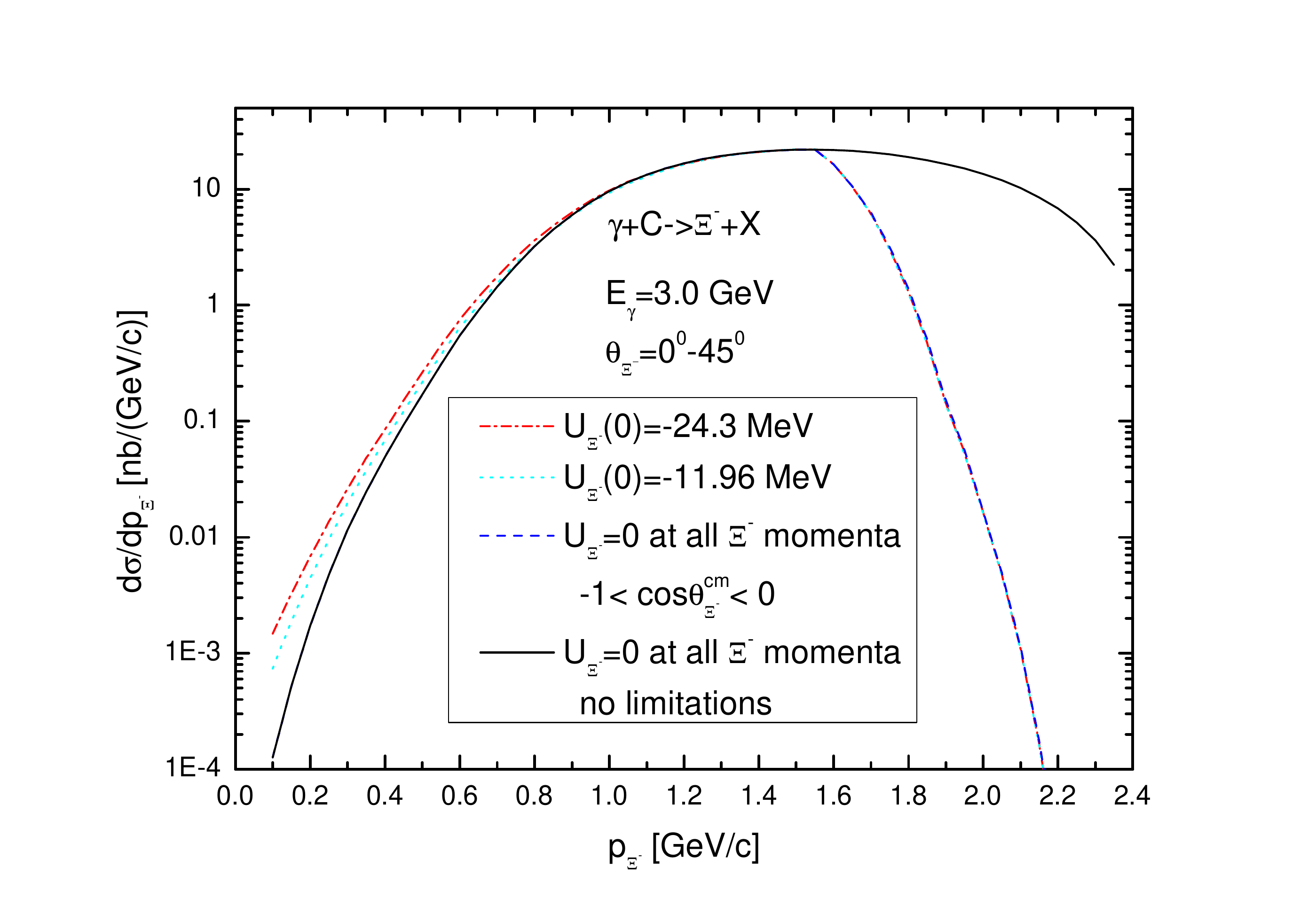}
\vspace*{-2mm} \caption{(Color online) Momentum differential cross sections for the production of $\Xi^-$
hyperons from the direct ${\gamma}p \to {K^+}{K^+}{\Xi^-}$ and ${\gamma}n \to {K^+}{K^0}{\Xi^-}$
processes in the laboratory polar angular range of 0$^{\circ}$--45$^{\circ}$ in the interaction of photons
of energy of 3.0 GeV with $^{12}$C nucleus. They were calculated for two different momentum dependences of
the $\Xi^-$ hyperon effective scalar potential $U_{\Xi^-}$ at density $\rho_0$ with the values
$U_{\Xi^-}(0)=-24.3$ MeV and $U_{\Xi^-}(0)=-11.96$ MeV, presented in Fig. 1, for zero potential at all $\Xi^-$ momenta, requiring that the $\Xi^-$ hyperons go backwards in the center-of-mass system
of the incident photon beam and a target nucleon at rest, as well as for zero potential at all $\Xi^-$ momenta
without any constraints on the $\Xi^-$
emission angle in this system.}
\label{void}
\end{center}
\end{figure}
\begin{figure}[!h]
\begin{center}
\includegraphics[width=18.0cm]{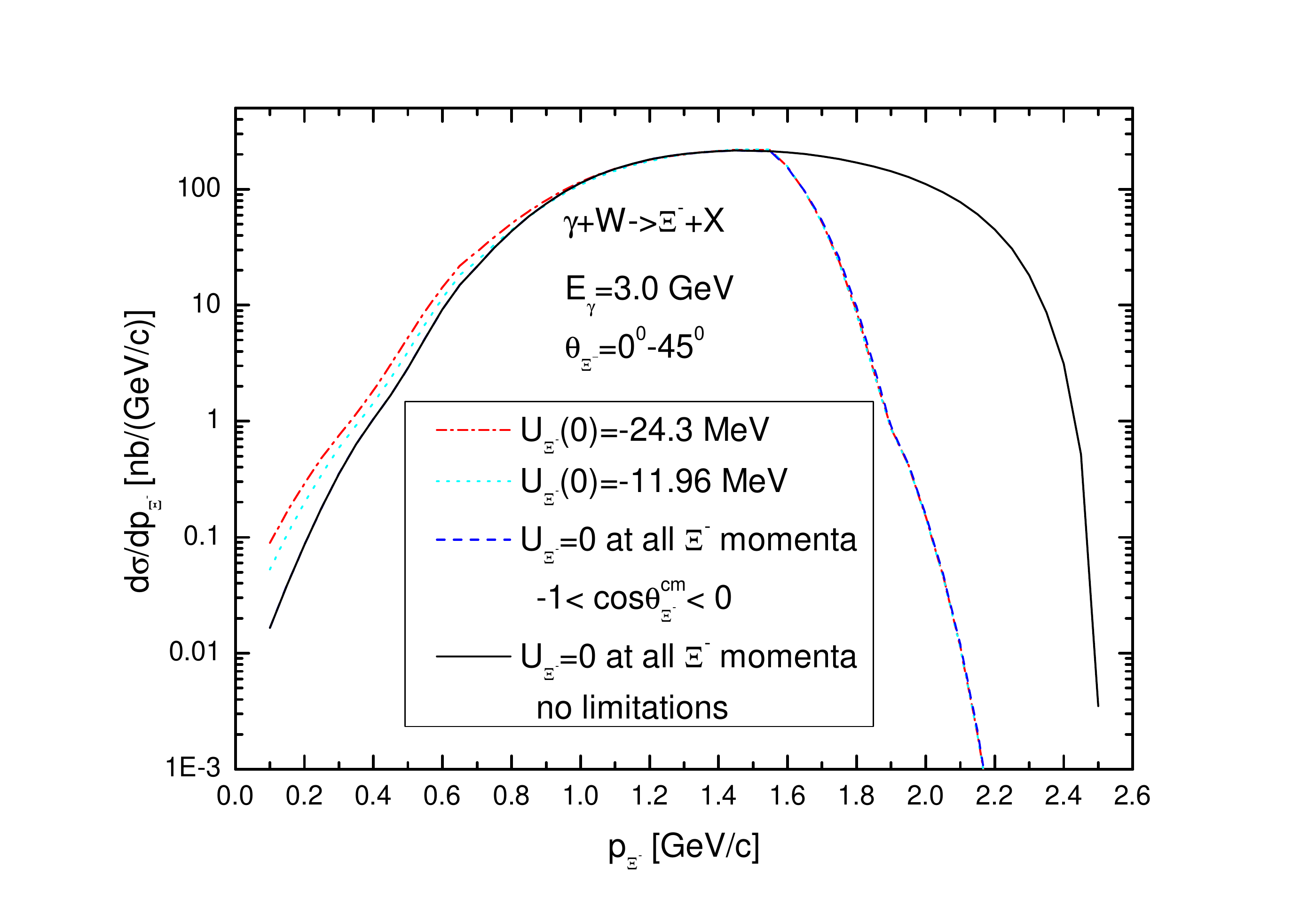}
\vspace*{-2mm} \caption{(Color online) The same as in Fig. 5,
but for the W target nucleus.}
\label{void}
\end{center}
\end{figure}
\begin{figure}[!h]
\begin{center}
\includegraphics[width=18.0cm]{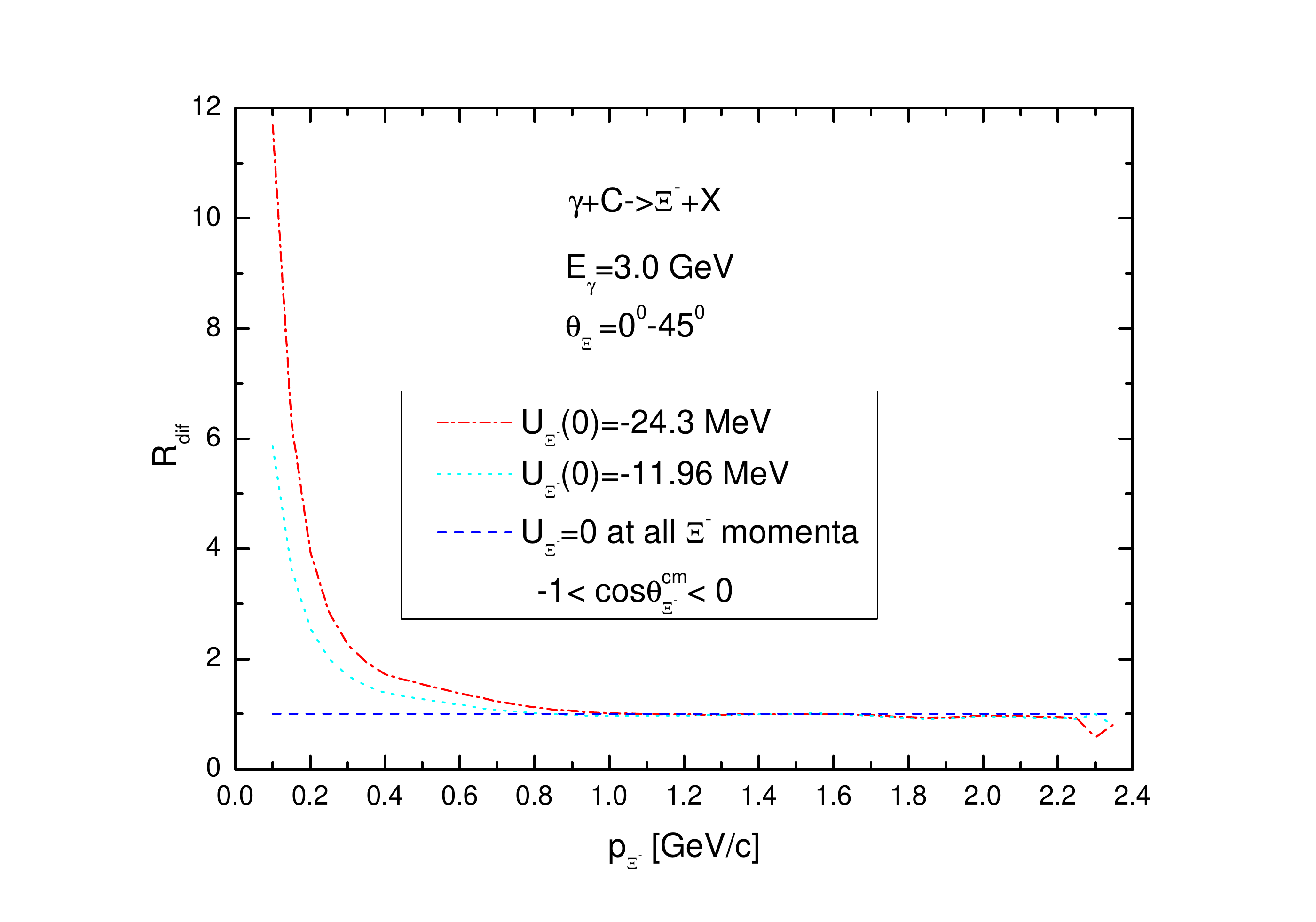}
\vspace*{-2mm} \caption{(Color online) Ratio between the differential cross sections for $\Xi^-$
production on $^{12}$C target nucleus in the angular region of 0$^{\circ}$--45$^{\circ}$ by photons
of energy of 3.0 GeV, calculated for two different momentum dependences of
the $\Xi^-$ hyperon effective scalar potential $U_{\Xi^-}$ at density $\rho_0$ with the values
$U_{\Xi^-}(0)=-24.3$ MeV and $U_{\Xi^-}(0)=-11.96$ MeV and presented in Fig. 1 as well as for zero potential at all $\Xi^-$ momenta and without this potential, requiring that the $\Xi^-$ hyperons go backwards in the center-of-mass system of the incident photon beam and a target nucleon at rest, as a function of $\Xi^-$ momentum.}
\label{void}
\end{center}
\end{figure}
\begin{figure}[!h]
\begin{center}
\includegraphics[width=18.0cm]{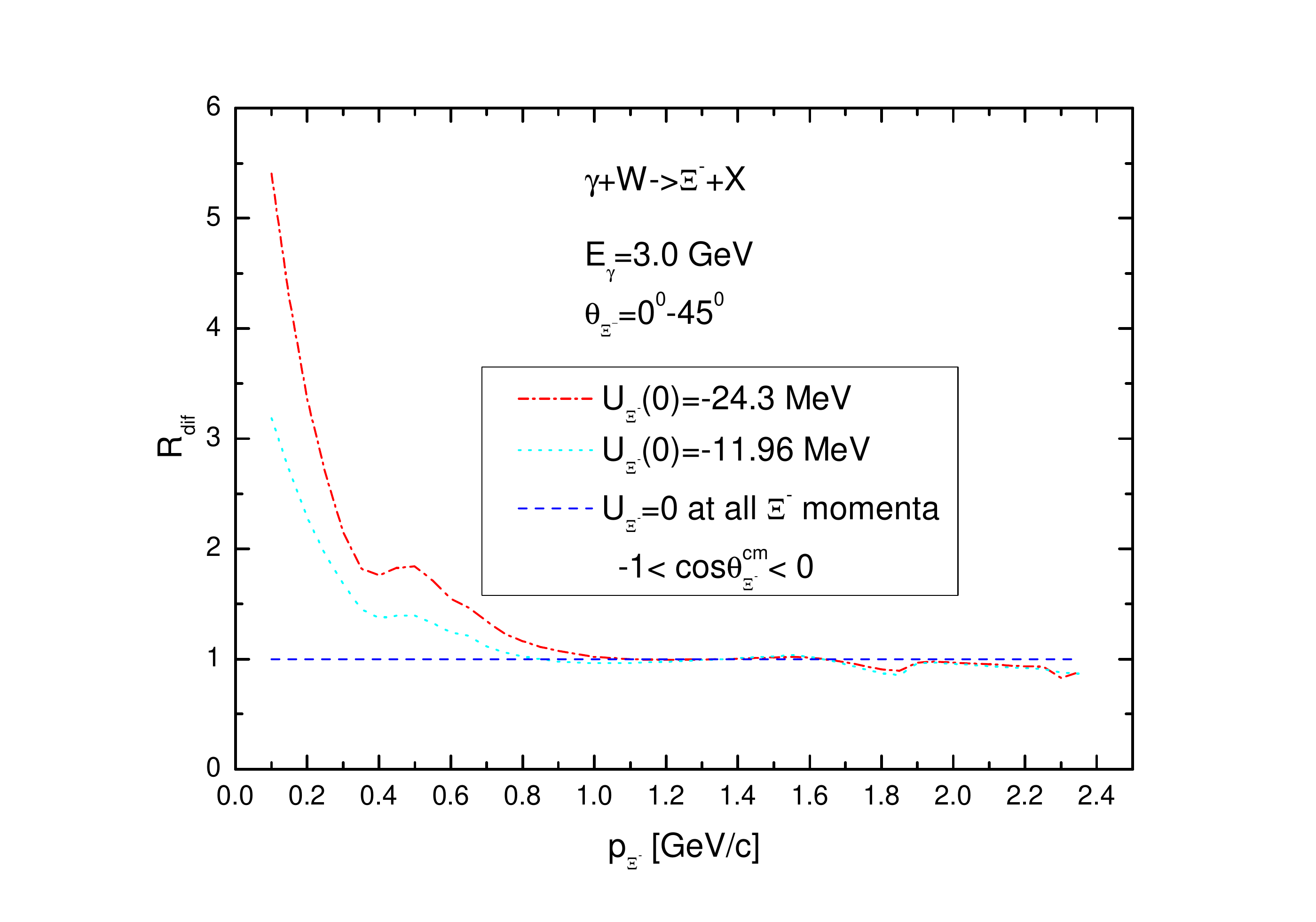}
\vspace*{-2mm} \caption{(Color online) The same as in Fig. 7,
but for the W target nucleus.}
\label{void}
\end{center}
\end{figure}
\begin{figure}[!h]
\begin{center}
\includegraphics[width=18.0cm]{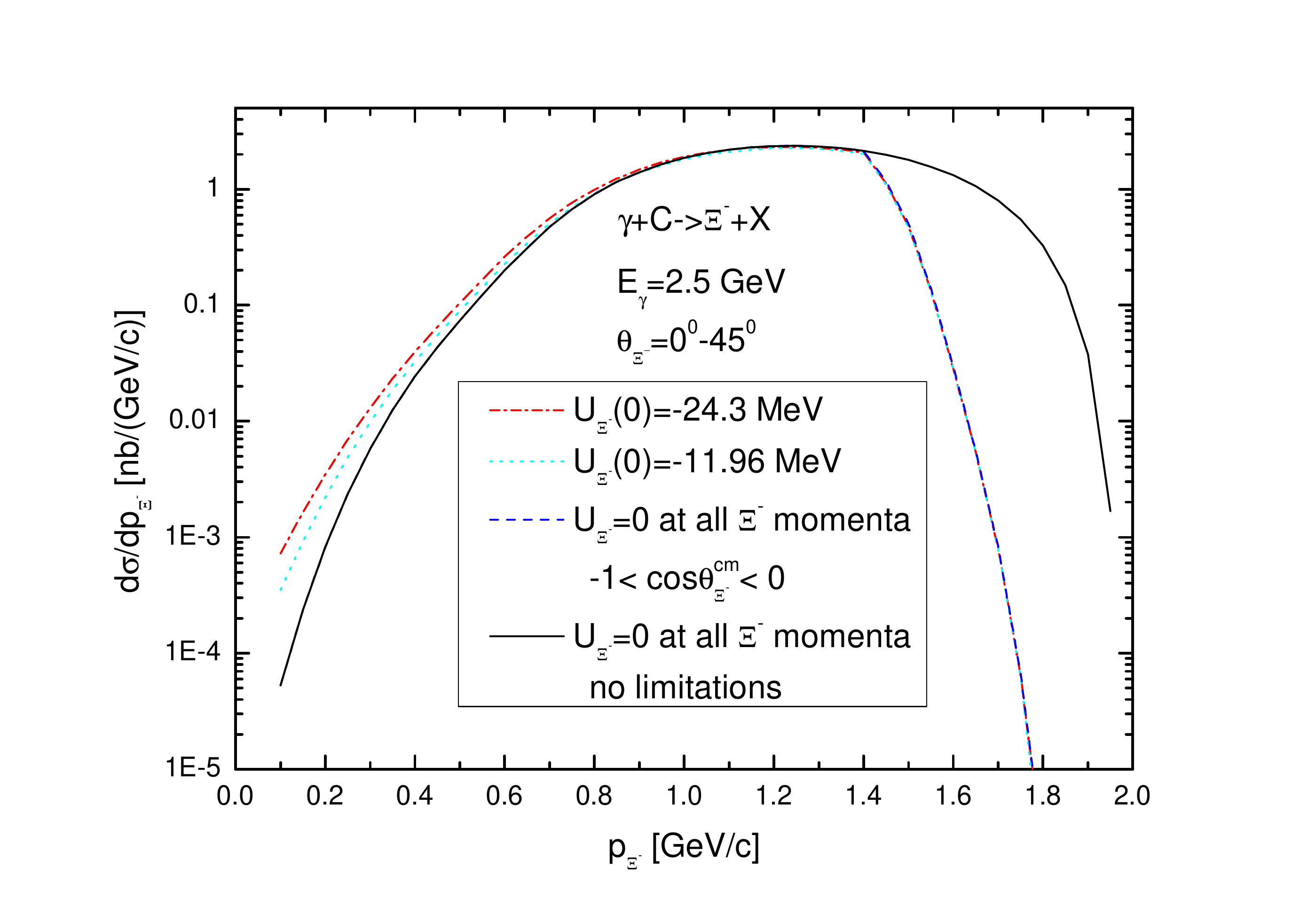}
\vspace*{-2mm} \caption{(Color online) Momentum differential cross sections for the production of $\Xi^-$
hyperons from the direct ${\gamma}p \to {K^+}{K^+}{\Xi^-}$ and ${\gamma}n \to {K^+}{K^0}{\Xi^-}$
processes in the laboratory polar angular range of 0$^{\circ}$--45$^{\circ}$ in the interaction of photons
of energy of 2.5 GeV with $^{12}$C nucleus. They were calculated for two different momentum dependences of
the $\Xi^-$ hyperon effective scalar potential $U_{\Xi^-}$ at density $\rho_0$ with the values
$U_{\Xi^-}(0)=-24.3$ MeV and $U_{\Xi^-}(0)=-11.96$ MeV, presented in Fig. 1, for zero potential at all $\Xi^-$ momenta, requiring that the $\Xi^-$ hyperons go backwards in the center-of-mass system
of the incident photon beam and a target nucleon at rest, as well as for zero potential at all $\Xi^-$ momenta
without any constraints on the $\Xi^-$ emission angle in this system.}
\label{void}
\end{center}
\end{figure}
\begin{figure}[!h]
\begin{center}
\includegraphics[width=18.0cm]{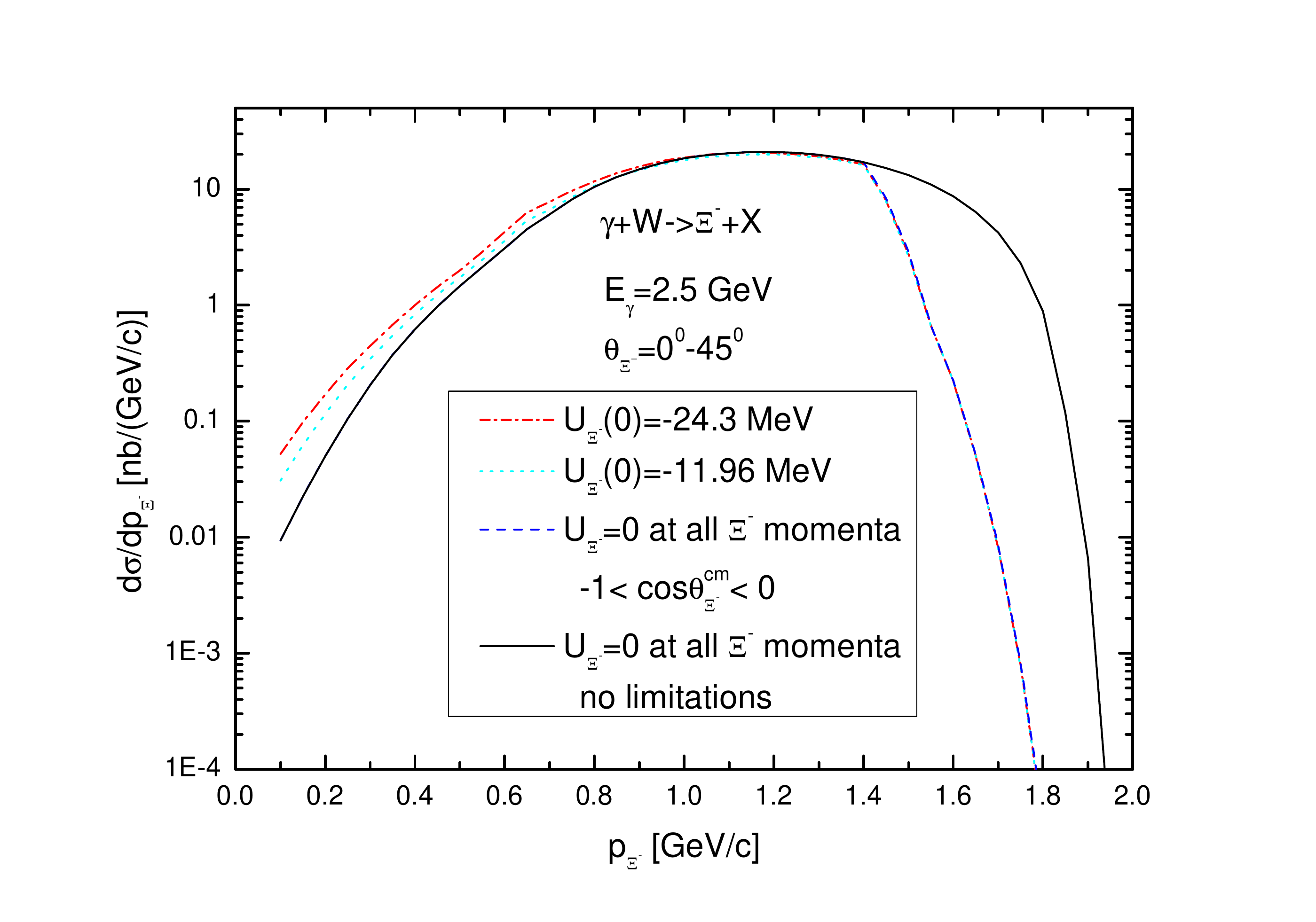}
\vspace*{-2mm} \caption{(Color online) The same as in Fig. 9,
but for the W target nucleus.}
\label{void}
\end{center}
\end{figure}

  Before closing this section, let us more closely consider, using the relativistic kinematics,
more simpler case of the production of $\Xi^-$ hyperons
in the elementary reaction ${\gamma}p \to K^+K^+\Xi^-$ proceeding on a free target proton being at rest.
This will help us to get some feeling about their kinematic characteristics allowed in this reaction at incident
photon energies of interest. The differential cross section
$d\sigma_{{\gamma}p \to K^+{K^+}{\Xi^-}}[\sqrt{s},m_{K^+},m_{K^+},m_{\Xi^-},{\bf p}_{\Xi^-}]/d{\bf p}_{\Xi^-}$ 
of this reaction can be obtained from more general one (18) in the limits: ${\bf p}_t \to 0$,
$E_t \to m_p$, $<m^*_{K^+}> \to m_{K^+}$, $<m^*_{\Xi^-}> \to m_{\Xi^-}$,
${\bf p}^{\prime}_{\Xi^-} \to {\bf p}_{\Xi^-}$, $E^{\prime}_{\Xi^-} \to E_{\Xi^-}$, $E^*_{\gamma} \to E_{\gamma}$,
$E^{*{\rm th}}_{\gamma} \to E^{\rm th}_{\gamma}$, $s^* \to s$ and $s^*_{K^+K^+} \to s_{K^+K^+}$.
The quantity $s_{K^+K^+}$, entering into it, must be greater than or equal to 4$m^2_{K^+}$. When the velocity
$v_{\rm cm}$ of the ${\gamma}p$ center-of-mass system in the laboratory frame is greater than or equal to the maximal $\Xi^-$ hyperon velocity $v^{*{\rm max}}_{\Xi^-}$ in this system ($v_{\rm cm} \ge v^{*{\rm max}}_{\Xi^-}$), the  condition $s_{K^+K^+} \ge 4m^2_{K^+}$ leads to the
fact that the polar $\Xi^-$ production angle $\theta_{\Xi^-}$ in this frame ranges from 0 to a maximal value
$\theta^{\rm max}_{\Xi^-}={\rm arcsin}[p^{*{\rm max}}_{\Xi^-}/(\gamma_{\rm cm}v_{\rm cm}m_{\Xi^-})]$.
Taking into account that the velocity $v^{*{\rm max}}_{\Xi^-}$ can be expressed via maximal $\Xi^-$ hyperon
momentum $p^{*{\rm max}}_{\Xi^-}$ and its maximal total energy $E^{*{\rm max}}_{\Xi^-}$ in the ${\gamma}p$ c.m.s.
as
\begin{equation}
v^{*{\rm max}}_{\Xi^-}=p^{*{\rm max}}_{\Xi^-}/E^{*{\rm max}}_{\Xi^-},
\end{equation}
where
\begin{equation}
p^{*{\rm max}}_{\Xi^-}=\frac{1}{2\sqrt{s}}\lambda(s,m^2_{\Xi^-},4m^2_{K^+}),\,\,\,
E^{*{\rm max}}_{\Xi^-}=\sqrt{(p^{*{\rm max}}_{\Xi^-})^2+m^2_{\Xi^-}},
\end{equation}
we get, for instance, that $v_{\rm cm}=0.762$, $v^{*{\rm max}}_{\Xi^-}=0.377$ and, respectively,
$\theta^{\rm max}_{\Xi^-}=20.3^{\circ}$ at $E_{\gamma}=3.0$ GeV (cf. footnote 7) given above).
At this energy, the value of the $\Xi^-$ hyperon total energy $E_{\Xi^-}(\theta^{\rm max}_{\Xi^-})$ in the l.s.,
corresponding to its maximal production angle $\theta^{\rm max}_{\Xi^-}=20.3^{\circ}$ is equal to
$E_{\Xi^-}(\theta^{\rm max}_{\Xi^-})={\gamma_{\rm cm}}m^2_{\Xi^-}/E^{*{\rm max}}_{\Xi^-}=1.89$ GeV.
This energy corresponds to the $\Xi^-$ momentum $p_{\Xi^-}(\theta^{\rm max}_{\Xi^-})=1.35$ GeV/c (cf. Fig. 3).
In the case when $v_{\rm cm} \ge v^{*{\rm max}}_{\Xi^-}$, as also follows from the condition
$s_{K^+K^+} \ge 4m^2_{K^+}$, the laboratory $\Xi^-$ hyperon momentum $p_{\Xi^-}$ at given production angle
$\theta_{\Xi^-}$, belonging to the interval
$[0,{\rm arcsin}[p^{*{\rm max}}_{\Xi^-}/(\gamma_{\rm cm}v_{\rm cm}m_{\Xi^-})]]$, varies within the range:
\begin{equation}
\frac{p_{\gamma}\sqrt{s}E^{*{\rm max}}_{\Xi^-}\cos{\theta_{\Xi^-}}-(E_{\gamma}+m_p)\sqrt{s}\sqrt{(p^{*{\rm max}}_{\Xi^-})^2-{\gamma^2_{\rm cm}}{v^2_{\rm cm}}m^2_{\Xi^-}\sin^2{\theta_{\Xi^-}}}}{(E_{\gamma}+m_p)^2-p^2_{\gamma}\cos^2{\theta_{\Xi^-}}} \le
\end{equation}
$$
p_{\Xi^-} \le
\frac{p_{\gamma}\sqrt{s}E^{*{\rm max}}_{\Xi^-}\cos{\theta_{\Xi^-}}+(E_{\gamma}+m_p)\sqrt{s}\sqrt{(p^{*{\rm max}}_{\Xi^-})^2-{\gamma^2_{\rm cm}}{v^2_{\rm cm}}m^2_{\Xi^-}\sin^2{\theta_{\Xi^-}}}}{(E_{\gamma}+m_p)^2-p^2_{\gamma}\cos^2{\theta_{\Xi^-}}}.
$$
In the case of zero $\Xi^-$ production angle the expression (39) is reduced to the following more simpler form:
\begin{equation}
{\gamma_{\rm cm}}E^{*{\rm max}}_{\Xi^-}(v_{\rm cm}-v^{*{\rm max}}_{\Xi^-}) \le p_{\Xi^-} \le
{\gamma_{\rm cm}}E^{*{\rm max}}_{\Xi^-}(v_{\rm cm}+v^{*{\rm max}}_{\Xi^-}).
\end{equation}
The results of our calculations by (39), (40) for the kinematically allowed $\Xi^-$ momenta in the
laboratory system for the $\Xi^-$ production angles of 0, 5, 10, 15, 20 and 20.25$^{\circ}$ at
$E_{\gamma}=3.0$ GeV are given in Fig. 3. It is seen that, as the $\Xi^-$ production angle increases,
the kinematically allowed $\Xi^-$ momentum range becomes narrower. Thus, for example, the laboratory
$\Xi^-$ hyperon momenta belong to the ranges of $0.85~{\rm GeV/c} \le p_{\Xi^-} \le 2.51~{\rm GeV/c}$,
$0.98~{\rm GeV/c} \le p_{\Xi^-} \le~1.99~{\rm GeV/c}$ and
$1.33~{\rm GeV/c} \le p_{\Xi^-} \le~1.37~{\rm GeV/c}$ at its production angles of 0, 15 and 20.25$^{\circ}$,
respectively. The maximal momentum range is at zero $\Xi^-$ production angle. Visual inspection of Fig. 3
tells us that for any given $\Xi^-$ hyperon momentum $p_{\Xi^-}$ from this range its laboratory production
angle $\theta_{\Xi^-}$ varies within the limits:
\begin{equation}
0 \le \theta_{\Xi^-} \le \theta^{\rm max}_{\Xi^-}(p_{\Xi^-}) \le \theta^{\rm max}_{\Xi^-},
\end{equation}
which corresponds to
\begin{equation}
\cos{\theta^{\rm max}_{\Xi^-}} \le \cos{\theta^{\rm max}_{\Xi^-}(p_{\Xi^-})} \le \cos{\theta_{\Xi^-}} \le 1.
\end{equation}
The kinematical considerations show that
\begin{equation}
\cos{\theta^{\rm max}_{\Xi^-}(p_{\Xi^-})}=\left({\gamma_{\rm cm}}E_{\Xi^-}-E^{*{\rm max}}_{\Xi^-}\right)/
({\gamma_{\rm cm}}{v_{\rm cm}}p_{\Xi^-}).
\end{equation}

 It is further interesting to calculate the $\Xi^-$ momentum spectrum from the considered reaction
${\gamma}p \to K^+K^+\Xi^-$ in the cases when $\Xi^-$ hyperons go only backwards and also backwards plus
forwards in the ${\gamma}p$ c.m.s. When the $\Xi^-$ momenta belong to the interval (40), we define this
spectrum, in line with Eq. (24), as:
\begin{equation}
\frac{d\sigma_{{\gamma}p\to K^+K^+{\Xi^-}}
(E_{\gamma},p_{\Xi^-})}{dp_{\Xi^-}}=2\pi
\int\limits_{\beta}^{1}d\cos{\theta_{\Xi^-}}p^2_{\Xi^-}
\frac{d\sigma_{{\gamma}p\to K^+K^+{\Xi^-}}[\sqrt{s},m_{K^+},m_{K^+},m_{\Xi^-},{\bf p}_{\Xi^-}]}{d{\bf p}_{\Xi^-}}
Q(\cos{\theta^{\rm cm}_{\Xi^-}}),
\end{equation}
where $\beta=\cos{\theta^{\rm max}_{\Xi^-}}(p_{\Xi^-})$ as well as in the former case the "$\Xi^-$ phase space
eliminating" factor $Q$ is defined above by the Eq. (25) and in the latter case it is equal to 1. The results
of calculations based on the Eq. (44) at $E_{\gamma}=3.0$ GeV are presented in Fig. 4. It can be easily seen that
when this factor is inroduced, i.e. when only the backward c.m.s. $\Xi^-$ momenta are allowed, the high-momentum
tail of the full laboratory $\Xi^-$ spectrum is suppressed and now the achievable $\Xi^-$ momentum ranges from
relatively high value of 0.85 GeV/c (cf. Fig. 3) to a sufficiently high one of $\approx$ 1.8 GeV/c. At such high
$\Xi^-$ momentum values it is difficult to expect the strong sensitivity of the $\Xi^-$ hyperon yield from nuclei
to its in-medium modifications. But if the Fermi motion and binding of the struck target proton are taken into
account, a low-momentum tail of the $\Xi^-$ momentum distribution from the considered elementary reaction appears.
And this tail possesses already a definite sensitivity to these modifications (see below).
\begin{figure}[!h]
\begin{center}
\includegraphics[width=18.0cm]{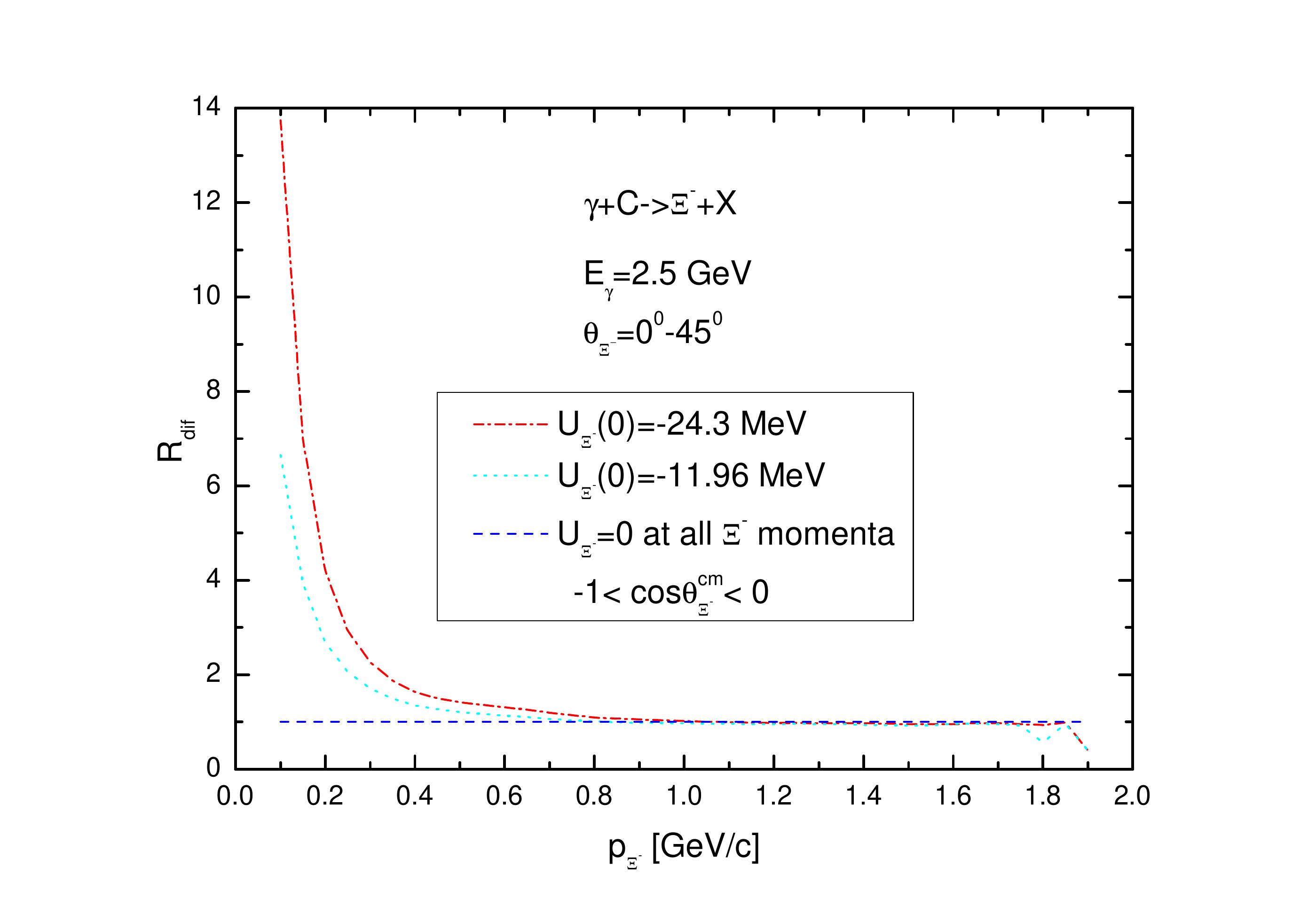}
\vspace*{-2mm} \caption{(Color online) Ratio between the differential cross sections for $\Xi^-$
production on $^{12}$C target nucleus in the angular region of 0$^{\circ}$--45$^{\circ}$ by photons
of energy of 2.5 GeV, calculated for two different momentum dependences of
the $\Xi^-$ hyperon effective scalar potential $U_{\Xi^-}$ at density $\rho_0$ with the values
$U_{\Xi^-}(0)=-24.3$ MeV and $U_{\Xi^-}(0)=-11.96$ MeV and presented in Fig. 1 as well as for zero potential at all $\Xi^-$ momenta and without this potential, requiring that the $\Xi^-$ hyperons go backwards in the center-of-mass system of the incident photon beam and a target nucleon at rest, as a function of $\Xi^-$ momentum.}
\label{void}
\end{center}
\end{figure}
\begin{figure}[!h]
\begin{center}
\includegraphics[width=18.0cm]{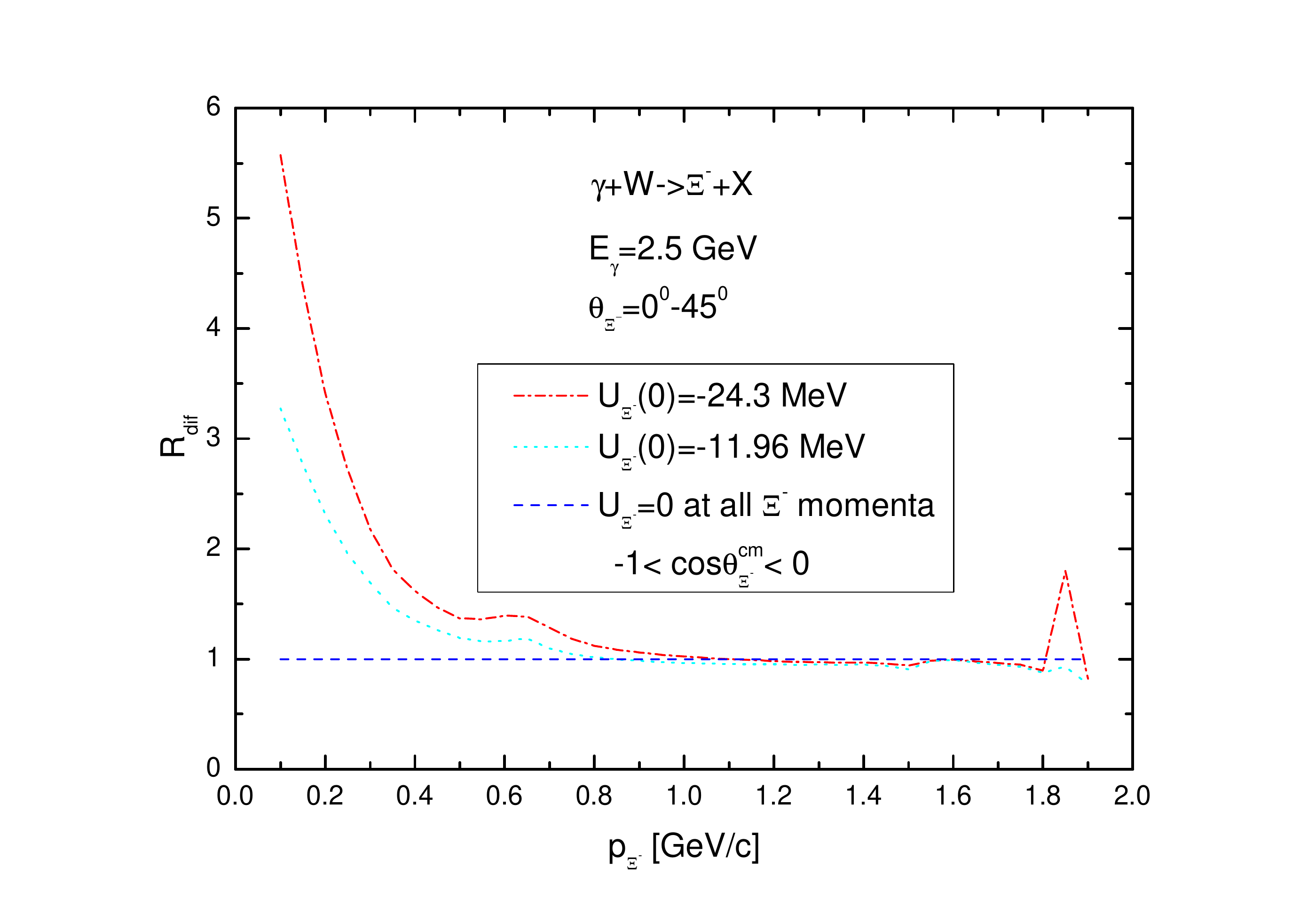}
\vspace*{-2mm} \caption{(Color online) The same as in Fig. 11,
but for the W target nucleus.}
\label{void}
\end{center}
\end{figure}
\begin{figure}[!h]
\begin{center}
\includegraphics[width=18.0cm]{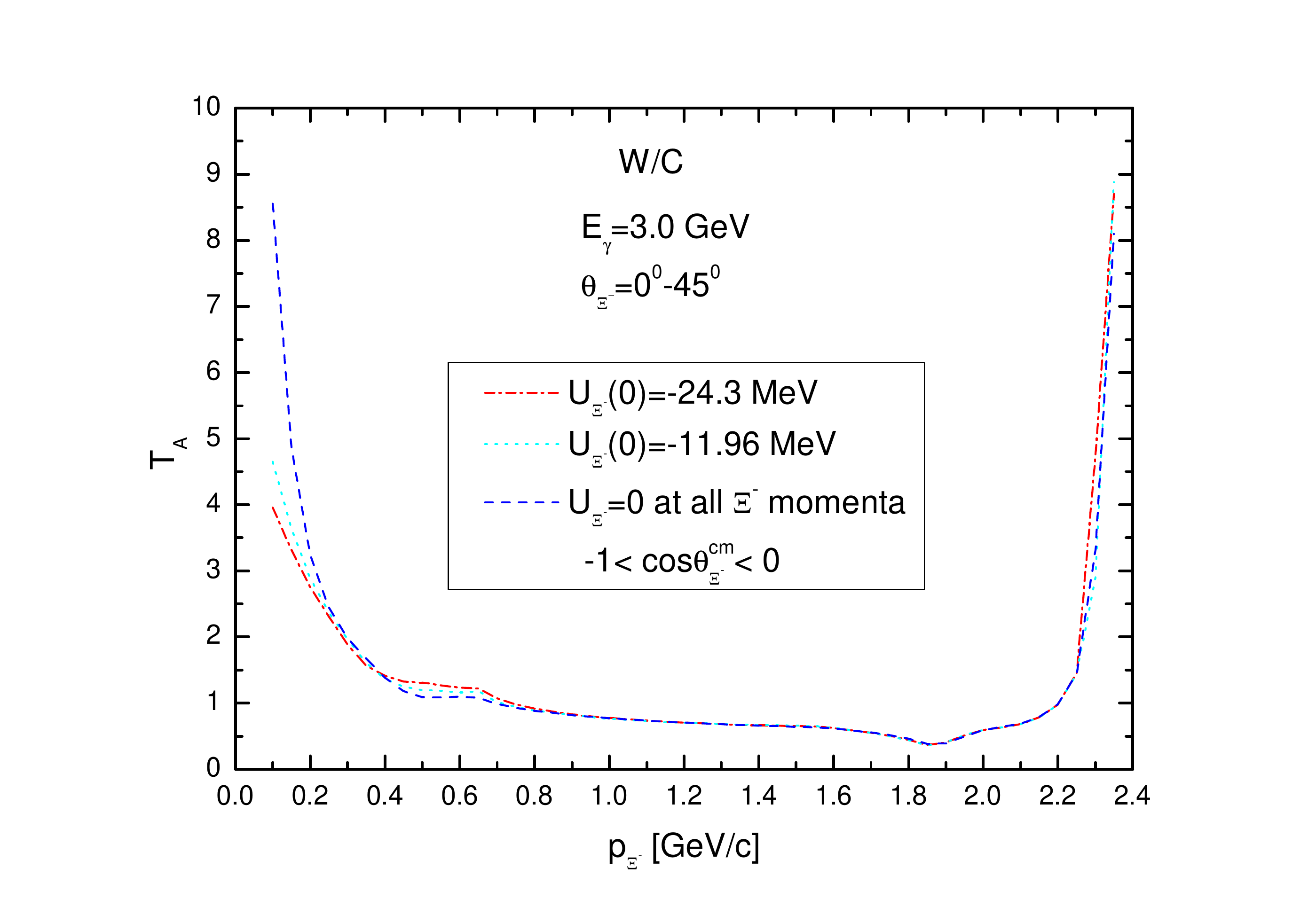}
\vspace*{-2mm} \caption{(Color online) Transparency ratio $T_A$ as a function of the $\Xi^-$  hyperon momentum
for combination $^{184}$W/$^{12}$C as well as for the $\Xi^-$ laboratory polar angular range of
0$^{\circ}$--45$^{\circ}$, for an incident photon energy of 3.0 GeV, calculated for two different momentum
dependences of the $\Xi^-$ hyperon effective scalar potential $U_{\Xi^-}$ at density $\rho_0$ with the values
$U_{\Xi^-}(0)=-24.3$ MeV and $U_{\Xi^-}(0)=-11.96$ MeV and presented in Fig. 1 as well as for zero potential at all $\Xi^-$ momenta, requiring that the $\Xi^-$ hyperons go backwards in the center-of-mass system of the incident photon beam and a target nucleon at rest.}
\label{void}
\end{center}
\end{figure}
\begin{figure}[!h]
\begin{center}
\includegraphics[width=18.0cm]{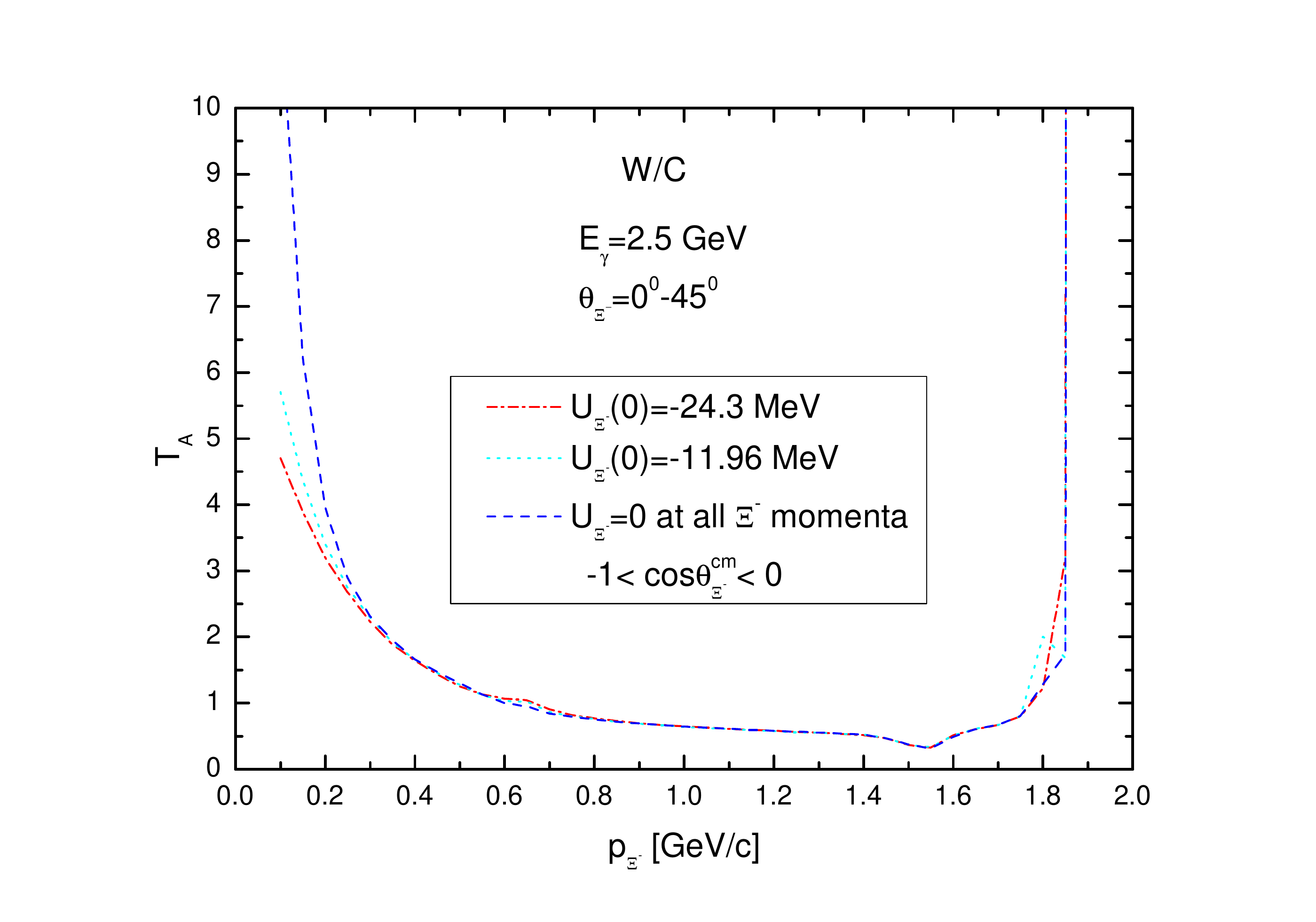}
\vspace*{-2mm} \caption{(Color online) The same as in Fig. 13,
but for the initial photon energy of 2.5 GeV.}
\label{void}
\end{center}
\end{figure}

\section*{3 Results}

\hspace{0.5cm} The model described above makes it possible to calculate the absolute values of the differential
cross sections for $\Xi^-$ photoproduction off nuclei and their ratios.
In the beginning, we consider the momentum dependences of these
cross sections from the production channels (1) and (2) in $\gamma$$^{12}$C and $\gamma$$^{184}$W
reactions. They were calculated in line with Eq. (24) for incident photon energy of 3.0 GeV
within three employed scenarios for the $\Xi^-$ effective scalar potential $U_{\Xi^-}$ at density $\rho_0$ for laboratory polar angles of 0$^{\circ}$--45$^{\circ}$ with and without ("no limitations" case)
imposing the phase space constraints on the $\Xi^-$ emission angle in the ${\gamma}p$ c.m. system with a target
proton being at rest.
These dependences are shown, respectively, in Figs. 5 and 6. One can see from them
that both the full $\Xi^-$ hyperon momentum distributions (solid curves) and modified ones (dotted-dashed, dotted, dashed lines) possess a distinct sensitivity to changes in this potential in the low-momentum region
of 0.1--0.8 GeV/c appearing, as was noted above, due to the binding of target nucleons and their Fermi motion
for both target nuclei. Whereas, at higher $\Xi^-$ momenta the impact of the $\Xi^-$ potential on these distributions
is negligible. In the low-momentum region, the differences between calculations corresponding to
different options for the $\Xi^-$ nuclear potential $U_{\Xi^-}$ are well distinguishable and experimentally
measurable. They are similar for both target nuclei at the considered photon energy of 3.0 GeV.
Thus, for instance, for vacuum $\Xi^-$ hyperon momenta of 0.2, 0.4, 0.6 GeV/c the inclusion of the
$\Xi^-$ momentum-dependent potential given above in Fig. 1 and having value of
$U_{\Xi^-}(0)=-11.96$ MeV at density $\rho_0$ leads in the case of $^{12}$C target nucleus
to an enhancement of the $\Xi^-$ production cross sections by factors of about 2.6, 1.4, 1.2, respectively,
as compared to those obtained for these momenta in the scenario when potential $U_{\Xi^-}=0$ MeV at all outgoing momenta. For the $^{184}$W nucleus these enhancement factors are about 2.3, 1.4, 1.2.
The inclusion of the $\Xi^-$ attractive momentum-dependent potential with value of $U_{\Xi^-}(0)=-24.3$ MeV
at density $\rho_0$ results in the further enhancement of the $\Xi^-$ hyperon production
cross sections by factors of about 1.5, 1.2, 1.2 and 1.5, 1.3, 1.3 compared to those calculated for the
above-mentioned $\Xi^-$ momentum-dependent potential for the same outgoing vacuum $\Xi^-$ momenta
of 0.2, 0.4, 0.6 GeV/c in the cases of $^{12}$C and $^{184}$W target nuclei, respectively.
The observed sensitivity of the strength of the low-momentum part of the $\Xi^-$ spectrum on the scalar
$\Xi^-$ potential $U_{\Xi^-}$ can be exploited to deduce the momentum dependence of this potential from the
direct comparison of the shapes of the calculated $\Xi^-$ hyperon differential distributions with that which
could be determined in the dedicated experiment in the present experimental facilities (for example, at the
CEBAF facility). Since the $\Xi^-$ hyperon production differential cross sections on $^{12}$C target nucleus
are less than those on the $^{184}$W by about of one to two orders of magnitude at all $\Xi^-$ momenta,
their measurements on a heavy nuclear targets like a tungsten one look quite promising and favorable.
At the photon beam energy of 3.0 GeV and with imposing the considered kinematical constraints on the $\Xi^-$
emission angle in the respective ${\gamma}p$ c.m.s. such measurements could be performed in a more narrower
$\Xi^-$ momentum range compared to that, corresponding to the case when these constraints are ignored, for
reliable determination of the $\Xi^-$ nuclear potential.

The sensitivity of the differential distributions, given in Figs. 5 and 6,
to the $\Xi^-$ hyperon in-medium modification at the saturation density $\rho_0$ may be more clearly seen
in the momentum dependences of the ratios $R_{dif}$ of these distributions,
calculated at incident photon energy of 3.0 GeV
for two considered $\Xi^-$ momentum-dependent potentials $U_{\Xi^-}$ and for zero momentum-independent potential,
to the analogous cross section, determined assuming that $U_{\Xi^-}=0$ MeV at all $\Xi^-$ momenta, with
requiring that the $\Xi^-$ hyperons go backwards in the center-of-mass system of the incident photon beam
and a target nucleon at rest.
We show in Figs. 7 and 8 such dependences on a linear scale for $^{12}$C and $^{184}$W nuclei, respectively.
It should be emphasized that such relative observable is more preferred compared to that based on the
absolute differential momentum distributions for the purpose of obtaining the definite information on particle in-medium properties, since the theoretical uncertainties associated with the description of the particle
production on nuclei substantially cancel out in it. It is nicely seen from these figures that there are indeed
experimentally measurable and similar differences for the $\Xi^-$ momenta $\le$ 0.8 GeV/c
between the calculations corresponding to the adopted options for the scalar potential $U_{\Xi^-}$ for both
target nuclei. Whereas, at higher $\Xi^-$ momenta these differences are practically indistinguishable.

We, therefore, come to the conclusion  that the most interesting low-momentum in-medium properties
of $\Xi^-$ hyperons could be definitely studied via the momentum dependences of their absolute
(and relative) production cross sections in inclusive photon-induced reactions at incident photon
beam energy of $E_{\gamma}=3.0$ GeV.

The results of calculations of the $\Xi^-$ momentum distributions and their ratios at incident photon energy
of 2.5 GeV (which is only slightly above the $KK\Xi^-$ threshold), analogous to those presented above in
Figs. 5, 6 and 7, 8, are given in Figs. 9, 10 and 11, 12, respectively. It can be seen that the highest
sensitivity both the $\Xi^-$ differential cross sections and their ratios there is again in the low-momentum
region of 0.1--0.8 GeV/c and this sensitivity is practically similar to that observed above at beam energy
of 3.0 GeV. But at this energy the $\Xi^-$ production cross sections are greater than those, calculated at
incident energy of 2.5 GeV, by factors of about 2--4 in the low-momentum region. Therefore, they could be
measured in this region at energy $E_{\gamma}=3.0$ GeV to a substantially higher statistical accuracy.
This is in favor of the $\Xi^-$
momentum distribution measurements, allowing to shed light on the scalar $\Xi^-$-nucleus optical potential
at momenta below $\approx$ 1.0 GeV/c, at photon energies around 3.0 GeV.

 Now, we consider the effects from $\Xi^-$ in-medium modification on the momentum dependence
of the transparency ratio $T_A$ for $\Xi^-$ hyperons.
Figures 13 and 14 show this dependence for the W/C combination for $\Xi^-$ hyperons produced
in the elementary processes (1) and (2) at laboratory polar angles $\le$ 45$^{\circ}$
by 3.0 and 2.5 GeV photons, respectively. It is calculated in line with Eq. (30)
for the employed scenarios for the $\Xi^-$  hyperon effective scalar potential $U_{\Xi^-}$ at
normal nuclear matter density $\rho_0$
and for the $\Xi^-$ emission angle in the respective ${\gamma}p$ c.m.s.
It is seen that for both photon energies the sensitivity of the
transparency ratio $T_A$ to the considered variations in the nuclear potential $U_{\Xi^-}$ is low
practically at all outgoing $\Xi^-$ momenta. This means that the momentum dependence of the
transparency ratio $T_A$ for $\Xi^-$ hyperons can hardly be used to determine this potential reliably.

Thus, we come to the conclusion that the $\Xi^-$ differential
(absolute and relative) cross section measurements in near-threshold photon-induced reactions
on nuclear targets might allow one to shed light on the momentum dependence of the $\Xi^-$ hyperon
effective scalar potential in cold nuclear matter at density $\rho_0$ for momenta below $\approx$ 1.0 GeV/c.
However, the transparency ratio measurements for $\Xi^-$ hyperons in these reactions cannot serve as a reliable
tool for determining this dependence.

\section*{4 Conclusions}

\hspace{0.5cm} In this paper we study the near-threshold inclusive photoproduction of $\Xi^-$ hyperons
off $^{12}$C and $^{184}$W target nuclei within a first-collision model relying on the nuclear spectral function
and including incoherent $\Xi^-$ production in direct elementary ${\gamma}p \to K^+{K^+}\Xi^-$ and
${\gamma}n \to K^+{K^0}\Xi^-$ processes. The model takes into account
the impact of the nuclear effective scalar $K^+$, $K^0$, $\Xi^-$ and their Coulomb potentials on these
processes as well as the absorption of final $\Xi^-$ hyperons in nuclear matter, the binding of intranuclear
nucleons and their Fermi motion. We calculate the absolute differential cross sections and their ratios for the production of $\Xi^-$ hyperons off these nuclei at laboratory polar angles $\le$ 45$^{\circ}$ by photons
with energies of 2.5 and 3.0 GeV, with and without imposing the phase space constraints
on the $\Xi^-$ emission angle in the respective ${\gamma}p$ center-of-mass system.
We also calculate the momentum dependence of the transparency ratio for $\Xi^-$ hyperons for the
$^{184}$W/$^{12}$C combination at these photon energies.
We show that the $\Xi^-$ momentum distributions (absolute and relative)
at the adopted initial photon energies possess a definite sensitivity to the considered changes
in the scalar $\Xi^-$ nuclear mean-field potential at saturation density $\rho_0$ in the low-momentum range
of 0.1--0.8 GeV/c. This would permit evaluating the $\Xi^-$ potential in this momentum range
experimentally. We also demonstrate that the momentum dependence of the transparency ratio for $\Xi^-$
hyperons for both photon energies can hardly be used to determine this potential reliably.
Therefore, the precise measurements of the differential cross sections (absolute and relative)
for inclusive $\Xi^-$ hyperon photoproduction on nuclei near threshold in a dedicated experiment at
the CEBAF facility will provide valuable information on the $\Xi^-$ in-medium properties,
which will be supplementary to that inferred from studying of the ($K^-$,$K^+$) reactions
at initial momenta of 1.6--1.8 GeV/c.
\\
\\

\end{document}